\documentclass[pdflatex,sn-mathphys-num]{sn-jnl}

\usepackage{graphicx}%
\usepackage{multirow}%
\usepackage{amsmath,amssymb,amsfonts}%
\usepackage{amsthm}%
\usepackage{mathrsfs}%
\usepackage[title]{appendix}%
\usepackage{xcolor}%
\usepackage{textcomp}%
\usepackage{manyfoot}%
\usepackage{booktabs}%
\usepackage{algorithm}%
\usepackage{algorithmicx}%
\usepackage{algpseudocode}%
\usepackage{listings}%

\theoremstyle{thmstyleone}%
\theoremstyle{thmstyletwo}%

\theoremstyle{thmstylethree}%

\graphicspath{{./figures}}

\begin{document}

\title[Optimizing Neural Networks for Bermudan Option Pricing: Convergence Acceleration, Future Exposure Evaluation and Interpolation in Counterparty Credit Risk]{Optimizing Neural Networks for Bermudan Option Pricing: Convergence Acceleration, Future Exposure Evaluation and Interpolation in Counterparty Credit Risk}


\author*[1]{\fnm{Vikranth} \sur{Lokeshwar Dhandpani}}\email{vikranthl@iisc.ac.in}
\author*[1]{\fnm{Shashi} \sur{Jain}}\email{shashijain@iisc.ac.in}
\affil*[1]{\orgdiv{Department of Management Studies}, \orgname{Indian Institute of Science}, \orgaddress{\street{Malleshwara}, \city{Bangalore}, \postcode{560012}, \state{Karnataka}, \country{India}}}


\abstract{This paper presents a Monte-Carlo-based artificial neural network framework for pricing Bermudan options, offering several notable advantages. These advantages encompass the efficient static hedging of the target Bermudan option and the effective generation of exposure profiles for risk management. We also introduce a novel optimisation algorithm designed to expedite the convergence of the neural network framework proposed by \citeauthor{lokeshwar2022explainable} (\citeyear{lokeshwar2022explainable}) supported by a comprehensive error convergence analysis. We conduct an extensive comparative analysis of the Present Value (PV) distribution under Markovian and no-arbitrage assumptions. We compare the proposed neural network model in conjunction with the approach initially introduced by \citeauthor{longstaff2001valuing} (\citeyear{longstaff2001valuing}) and benchmark it against the COS model, the pricing model pioneered by \citeauthor{fang2009pricing} (\citeyear{fang2009pricing}), across all Bermudan exercise time points. Additionally, we evaluate exposure profiles, including Expected Exposure and Potential Future Exposure, generated by our proposed model and the Longstaff-Schwartz model, comparing them against the COS model. We also derive exposure profiles at finer non-standard grid points or risk horizons using the proposed approach, juxtaposed with the Longstaff Schwartz method with linear interpolation and benchmark against the COS method. In addition, we explore the effectiveness of various interpolation schemes within the context of the Longstaff-Schwartz method for generating exposures at finer grid horizons.}

\keywords{Bermudan Option Pricing, Artificial Neural Networks, Error Convergence, Exposure management}



\maketitle

\section{Introduction}

\noindent Financial markets have witnessed significant growth in the complexity of derivative instruments over the past few decades. Among these, Bermudan options are prominent due to their flexibility regarding exercise dates, which strikes a balance between European and American options. Bermudan options grant the holder the right to exercise at specific, predetermined dates until expiration. This unique feature introduces a layer of complexity into pricing and risk management, making them an intriguing subject of study for financial researchers and practitioners. This paper mainly addresses the critical challenges in pricing, hedging and Counterparty Credit Risk (CCR) management of Bermudan options using artificial neural networks.\\

\noindent The valuation of Bermudan options is fundamentally a dynamic programming problem, with the primary challenge in the computation of conditional expectations required for determining the continuation value. The research by \citeauthor{carriere1996valuation} (\citeyear{carriere1996valuation}) was one of the early works to approximate the conditional expectation under the Monte Carlo framework by non-parametric regression techniques (spline and local regression). The usage of Monte-Carlo backward induction-based regression methods to estimate the conditional expectation for pricing Bermudan options became widely popular after the work by \citeauthor{longstaff2001valuing} (\citeyear{longstaff2001valuing}) (called the Longstaff-Schwartz method in this paper). In this method, the key challenge is the choice of basis function, which is arbitrary and depends on factors like underlying risk factors, payoff structure and market environment. \citeauthor{tsitsiklis2001regression} (\citeyear{tsitsiklis2001regression}) adapted a parametric function of time-state pair instead of an independently parameterised value function at each exercise time. \citeauthor{clement2002analysis} (\citeyear{clement2002analysis}) used Monte-Carlo simulation coupled with least squares regression to compute the value function by approximating the conditional expectations by projections on a finite set of functions. \citeauthor{glasserman2004number} (\citeyear{glasserman2004number}) illustrated the convergence with polynomial basis functions under Brownian motion and geometric Brownian motion assumptions for the underlying. \\

\noindent Furthermore, research was focused on achieving both speed and accuracy in pricing early exercise options. An example of such research is the Fast Fourier method for early exercise options, as introduced by \citeauthor{lord2008fast} (\citeyear{lord2008fast}). This method, requiring only knowledge of the model's characteristic function, is particularly suitable for exponential Levy models, including those featuring exponentially affine jump-diffusion characteristics. One of the key contributions of \citeauthor{fang2009pricing} (\citeyear{fang2009pricing}) includes a pricing method (called COS method in this paper) based on Fourier-cosine expansions for early exercise options, which works well for exponential Levy asset price models. This work extends the novel pricing method initially introduced for European options in \citeauthor{fang2008novel} (\citeyear{fang2008novel}) to options with exotic features. \citeauthor{jain2015stochastic} (\citeyear{jain2015stochastic}) introduced a simulation-based stochastic grid bundling method under the multidimensional framework for pricing Bermudan options and fast approximation of the Greeks. With the advancement of computing capability, machine learning techniques became attractive solutions for pricing early exercise options. Some notable works include neural networks for American option pricing by \citeauthor{kohler2010pricing} (\citeyear{kohler2010pricing}), neural network approximation of conditional expectation within Longstaff and Schwartz algorithmic framework by \citeauthor{lapeyre2021neural} (\citeyear{lapeyre2021neural}) instead of standard least squares techniques, deep neural network framework to price American options by designing sequential neural networks to learn the difference of price functions between adjacent exercise time steps by \citeauthor{chen2021deep} (\citeyear{chen2021deep}), solution of high-dimensional optimal stopping problems using deep learning by \citeauthor{becker2021solving} (\citeyear{becker2021solving}). \\

\noindent After the 2007-2009 credit crisis, the Basel Committee on Banking Supervision (BCBS) significantly strengthened its Counterparty Credit Risk (CCR) framework. Specifically, the Internal Models Approach (IMA) involves estimating potential future exposure distributions of all transactions in a netting set at different time horizons and calculating key statistics like Expected Exposure (EE) profile (expected exposure across different risk horizons or time horizons), Potential Future Exposure (PFE) profiles for multiple requirements such as Credit Valuation Adjustment (CVA) calculation, setting limits, capital calculation and monitoring. EE is the expectation of the exposure distribution at a specified risk horizon, and the PFE represents a quantile (99\% in this paper) of the exposure distribution. Determining exposure profiles for early exercise options is challenging due to the computational complexity to calculate precise exposure across all the simulated points at each time horizon. \citeauthor{de2014efficient} (\citeyear{de2014efficient}) studied the exposure distributions by simulating risk factors using Monte Carlo simulation and evaluating the positions using the  COS method, finite difference method to solve partial differential equation, and stochastic grid bundling method. \citeauthor{karlsson2016counterparty} (\citeyear{karlsson2016counterparty}) extended the Stochastic Grid Bundling Method for the one-factor Gaussian short rate model to efficiently and accurately compute the EE, PFE and CVA for Bermudan swaptions. \citeauthor{gnoatto2023deep} (\citeyear{gnoatto2023deep}) introduced a coupled system of backward stochastic differential equations solved by recursive application of neural networks to calculate valuation adjustments. \\

\noindent \citeauthor{lokeshwar2022explainable} (\citeyear{lokeshwar2022explainable}) presented a generic regress-later Monte Carlo approach which uses neural networks to price multi-asset discretely-monitored contingent claims. The choice of neural network architecture provided a meaningful interpretation of the neural network in a  financial context. The interpretation demonstrated that any discretely monitored contingent claim (possibly high-dimensional and path-dependent) under Markovian and no-arbitrage assumptions can be semi-statically hedged using a portfolio of short-maturity options. For the one-dimensional case of Bermudan options, better convergence can be achieved and this paper focuses on that along with an angle of risk management. The key contributions are briefly outlined below:

\begin{itemize}
\item This paper concentrates on customizing the Regress-Later Neural Network (RLNN) introduced by \citeauthor{lokeshwar2022explainable} (\citeyear{lokeshwar2022explainable}) specifically for pricing \emph{single-asset} Bermudan options. We achieve faster convergence in finding optimal network parameters by adapting the algorithm for a single underlying risk factor. Our numerical examples showcase the proposed model's improved accuracy and quicker error convergence in pricing Bermudan options. We compare these results with the RLNN model, referencing values obtained through the COS method.\\

\item The paper also demonstrates the neural network's ability to generate future exposure distributions for Bermudan options at no additional cost, allowing the evaluation of exposure profiles, including EE and PFE. We compare exposure profiles from our model with those from the industry-standard Longstaff-Schwartz method (LSM). The COS method obtains reference values by pricing the target option for each simulated scenario on a future exposure date. \\

\item In the context of Counterparty Credit Risk (CCR), there's often a need to calculate future exposure at time grid points that don't coincide with the early exercise dates of Bermudan options. Typically, LSM regression for Bermudan option pricing occurs only at early exercise dates, necessitating additional regressions or interpolation to determine exposure at these grid points. This is particularly crucial for margined exposures with cash collateral, where exposure at the Margin Period of Risk may not align with the original regression dates. Our study demonstrates that our proposed model accurately computes exposures at these intermediate, non-standard grid points without performing additional regressions. We also present a comprehensive comparative analysis of exposures obtained using the LSM method at these time grid points. \\

\item The investigation also explores various interpolation methodologies that could be integrated into the LSM framework to generate these intermediate exposures. Subsequently, the potential interpolation schemes within the LSM framework are recommended and compared. This contribution sheds light on the intricacies of exposure generation and interpolation within the landscape of Counterparty Credit Risk.

\end{itemize}

\section{Problem Formulation}
\label{Problem Formulation}

\noindent This section introduces the notations\footnote{It is to be noted that $\cdot$ is generally used for multiplication, which includes matrix multiplication, and $\odot$ is used for element-wise multiplication between matrices or vectors of the same dimensions. In this paper, the function applied on each element of the vector (or matrix) is notationally mentioned as the function on the vector (or matrix), i.e., for any arbitrary vector $Q=(q_1, q_2, \ldots, q_d)^{\intercal} \in \mathbb{R}^{d}$, $(f(q_1), f(q_2), \ldots, f(q_d))^{\intercal}$ is recorded as $f(Q)$.} used and illustrates the Bermudan options pricing framework proposed in this paper.\\ 

\noindent We assume a complete probability space $(\Omega, \mathcal{F}, \mathbb{P})$,  filtration $\mathcal{F}_t: \ t \in [0,T]$ and an adapted underlying asset process $S_t$, $ \forall t$. The stochastic dynamics of the underlying asset are assumed to follow Geometric Brownian Motion (GBM), and therefore, 
 
\begin{equation}\label{GBMprocessOptimalRLNN}
S_{t} = S_{0} \cdot  exp \left(  \Big( \ r - \frac{\sigma^2}{2} \ \Big) \ t + \sigma \ Z_{t} \right) , 
\end{equation} 

\noindent where, $S_{0}$ is the initial value of the underlying at time $0$, $r$ is the constant risk-free interest rate, $\sigma$ is the constant volatility and $Z_{t}$ is Brownian Motion. \\

\noindent We aim to price the target Bermudan option, with strike $K$ $\in \mathbb{R}$, starting at time $t_0=0$ and expiring at $T$, with the right to exercise at each $t_{m}$, where, $m \in \{1, 2, 3, \ldots, M \}$, $t_{M} = T$ and $t_{m} - t_{m-1} = $ $\Delta t$  $\forall m$. \\ 

\noindent Let $h_t:= h(S_t)$ be an adapted process representing the option's intrinsic value; the holder of the option receives $max(h_t, 0)$ if the option is exercised at time $t$. $h_t = (S_t - K)$ for a call and $h_t = (K - S_t)$  for a put option. \\ 

Assuming a risk-free savings account process, $B_t = exp(r \cdot t)$, the Bermudan option price at $t_0$ is,
\begin{align}
\frac{V_{t_{0}} (S_{t_{0}})}{B_{t_{0}}} = \max_{\tau} \ \mathbb{E} \Bigg[\frac{h(S_{\tau})}{B_{\tau}}\Bigg] ,
\end{align}

\noindent where $V_{t}(.): \ t \in [0, \ T]$ is the option value function, and $\tau$ is the stopping time, taking values in the finite set $\{0, t_1, \ldots, T\}$. \\

\noindent The dynamic programming formulation to solve this optimisation problem is as follows. The value of the option at 
the terminal time $T$ is equal to the product's pay-off,
\begin{align}
V_T(S_T) = \max(h(S_T), 0).
\end{align}

\noindent Recursively, moving backwards in time, the following iteration is then solved, given $V_{t_m}$ has already been determined, the continuation or hold value $Q_{t_{m-1}}$ is given by:
\begin{align}
Q_{t_{m-1}} (S_{t_{m-1}}) = B_{t_{m-1}} \mathbb{E}\Bigg[  \frac{V_{t_m}(S_{t_m})}{B_{t_m}} \Bigg| S_{t_{m-1}}  \Bigg],
\end{align}

\noindent where, $\mathbb{E}$ stands for Expectation under the risk-neutral measure. \\

\noindent The Bermudan option value at time $t_{m-1}$ for the underlying state $S_{t_{m-1}}$ is then given by:
\begin{align}
V_{t_{m-1}} (S_{t_{m-1}}) = \max \big( h(S_{t_{m-1}}), Q_{t_{m-1}} (S_{t_{m-1}}) \big).
\end{align}

In the RLNN model by \citeauthor{lokeshwar2022explainable} (\citeyear{lokeshwar2022explainable}), the neural network $\tilde{G}^{\beta_{t_m}}$ was trained at each exercise time $t_m$ ($m>0$). The neural network inputs the simulated underlying price $S_{t_m}$ and outputs $\tilde{G}^{\beta_{t_m}} (S_{t_m})$, such that,  
\begin{align}
\beta_{t_m} &= \underset{\beta} {\mathrm{argmin}} \ \ ||V_{t_m}(S_{t_m}) - \tilde{G}^{\beta} (S_{t_m})||^{2},
\end{align}

\noindent where, $\beta_{t_m}$ are the optimal neural network parameters and $||.||$ is the L2-norm. \\

The conditional expectation of the neural network $\tilde{G}^{\beta_{t_m}}$ was then determined and utilised to predict the continuation value at $t_{m-1}$. The valuation process starts by setting up the neural network at $T$, where the option value is deterministically determined from the payoff function for a given underlying price. It is then recursively moved backwards at each exercise time to price the Bermudan option. This neural network is shortly referred to as RLNN. \\

\noindent This paper focuses on developing an enhanced neural network framework to predict the continuation value with faster convergence to price Bermudan options. A better parameter initialisation and a novel optimisation methodology are recommended to learn the neural network parameters $\beta_{t_m}$ for each $t_m$. 

\subsection{Neural Network Architecture}
\label{Neural Network Architecture}

We are interested in building the neural network $\tilde{G}^{\beta_{t_m}}$ at each exercise time $t_m$ with faster convergence. The neural network model inputs underlying asset price ($S_{t_m}$) and predicts target option value $V_{t_m}(S_{t_m})$. \\

We first see the interpretation of the proposed neural network to understand its architecture. The output of the neural network $\tilde{G}^{\beta_{t_m}} (S_{t_m})$ can be interpreted as the portfolio value (as of $t_m$) of $p$ European call and put options initiated at time $t_{m-1}$ and expiring at $t_m$. Let $W$ $\in$ $\mathbb{R}^{p}$ be the portfolio weights and $\boldsymbol{b}$ $\in$ $\mathbb{R}^{p}$ be the strikes of the constituent options. This portfolio is selected to replicate the target option at all simulated $S_{t_{m}}$ levels. Therefore, under no arbitrage, the portfolio inferred from each neural network $\tilde{G}^{\beta_{t_m}}$ will act as a replicating portfolio of the target option from $t_{m-1}$ till $t_{m}$, $\forall \ m $. The portfolio expires at each exercise point and can be re-invested to set up a new portfolio that replicates the target option until the next exercise date. Therefore, the neural network constructs a self-replicating static hedging portfolio for the target option between each exercise date. \\

\noindent With this interpretation in consideration, the neural network includes three layers:  

\begin{enumerate}

\item The first layer has a single node that inputs and fans out the underlying asset price to the hidden layer. 

\item The hidden layer has $p$ nodes with bias and Rectified Linear Unit (ReLU) activation function in each node. The bias term of each hidden node corresponds to the strike $b_i$ (where,  $i = 1, 2, \ldots, p$) of each constituent option in the static hedge portfolio and $\boldsymbol{b} = (b_1, b_2, \ldots, b_p)^{\intercal}$. There are two constraints induced in the neural network:  

\begin{itemize}
\item The neural network weights between the input and hidden layers are kept constant (with values $+1$ for the call and $-1$ for the put option nodes in the hidden layer).

\item The bias\footnote{In theory, we aim for the bias term $b_i$ to be greater than or equal to zero. However, in practice, we utilize the floor value $10^{-8}$ to mitigate potential numerical errors.} term is constrained such that $b_i$ $\ge 10^{-8}$ for all the hidden nodes. Additionally, the activation functions for the call and put option nodes correspond to their respective payoff functions, following ReLU structure.
\end{itemize}

\noindent Consequently, for any exercise time $t \in \{t_1, t_2, \ldots, t_M\}$, the output of each node is consistent with the payoff of a call or put option $\phi_{i_{cp}}(S_{t}, b_i) = max\big( i_{cp} \cdot (S_{t} - b_i), 0 \big)$, where, $i_{cp} = +1$ for call options and $-1$ for put options. Let $I_{cp} = (i_{cp})^{\intercal}_{i=1,2, \ldots, p} \in \mathbb{R}^{p}$ be call/put indicator vector or equivalently the constant neural network weights between the first layer and hidden layer. Therefore, the output payoff vector of the hidden layer can be represented as:

\begin{equation*}
 \phi(S_{t}, \boldsymbol{b}) = \big(\phi_{i_{cp}}(S_{t}, b_i) \big)_{i=1, \ldots, p}^{\intercal} \in \mathbb{R}^{p}
\end{equation*}

\item The third layer (output layer) has one node which inputs $W^{\intercal} \phi(S_{t}, \boldsymbol{b})$, where, $W = (w_1, w_2, ..., w_p)^{\intercal} \in \mathbb{R}^{p}$ are the neural network weights (between the hidden layer and output layer). The neural network weights $W$ correspond to the portfolio weights of the static hedge portfolio. Therefore, the output of the third layer corresponds to the payoff of the portfolio, as shown below: 

\begin{equation*}
\tilde{G}^{\beta_{t}} (S_{t}) = W^{\intercal} \phi(S_{t}, \boldsymbol{b}). 
\end{equation*}

\end{enumerate}

Therefore, the neural network parameters $\beta_{t_m}$ have two components - the strikes of the constituent options and portfolio weights:

\begin{align}
\beta_{t_m} := [W, \boldsymbol{b}] \ \ \ \forall m \in \{1, 2, \ldots, M\},
\end{align}

 The continuation value at time $t_{m-1}$ is the risk-neutral value of the portfolio inferred from parameters $\beta_{t_m}$. Each constituent option maturing in $\Delta t$ period can be valued at $t_{m-1}$ using the risk-neutral expectation of their payoffs as of $t_m$. Under the assumption that asset follows a GBM process, this expectation can be evaluated by the Black-Scholes pricing model by \citeauthor{black1973pricing} (\citeyear{black1973pricing}). In other words, the static hedge portfolio value at time $t_{m-1}$, denoted by $\mathbb{E}\left[\tilde{G}^{\beta_{t_m}} (S_{t_m}) \mid S_{t_{m-1}}\right]$, is the proposed continuation value $Q_{t_{m-1}} (S_{t_{m-1}})$. The option value $V_{t_{m-1}} (S_{t_{m-1}})$ at time $t_{m-1}$ is shown below.

\begin{align}\label{static-hedge-model-opt-value}
Q_{t_{m-1}} (S_{t_{m-1}}) &= \mathbb{E}\left[\tilde{G}^{\beta_{t_m}} (S_{t_m}) \mid S_{t_{m-1}}\right] , \nonumber \\
V_{t_{m-1}} (S_{t_{m-1}}) &= \max \ \Big(Q_{t_{m-1}} (S_{t_{m-1}}), h(S_{t_{m-1}})\Big),  \\
 & \ \ \ \ \ \ \ \ \forall m \in \{1, 2, \ldots, M\}, \nonumber \\ 
& \ \ \ \ \ \ \ \ V_{t_M}(S_{t_M}) = h(S_{t_M}). \nonumber
\end{align}

\noindent $V_{t_{0}} (S_{t_{0}})$ is the proposed time-zero price of the target Bermudan option by the static hedge model. This paper denotes the proposed neural network model as \emph{RLNN-OPT} (Optimal Regress-Later Neural Network). As the neural network portfolio serves as the recommended static hedge portfolio for the target option, it is also referred to interchangeably as the \emph{Static Hedge Model}.

\subsection{Design of Training data}
\label{Design of Training data}

\noindent The data for training the neural network is obtained by simulating $N$ paths of the underlying asset using the GBM process as defined in Equation \ref{GBMprocessOptimalRLNN} at all time points of interest $t = t_1, t_2, \ldots, t_{M}$. At each simulated path $\omega_j \in \Omega$ (where, $j = 1, 2, ....., N$) of the underlying $S_{t}$, the neural network is built as a function between underlying price as the independent variable and option value (as defined in Equation \ref{static-hedge-model-opt-value}) as the response variable. In other words, $\Bigg(S_t\big(\omega_j\big), V_{t} \big(S_t (\omega_j) \big)  \Bigg)_{j=1}^{N}$ is the feature variable and response variable data pair for training the neural network $ \forall t \in \{ t_1, t_2, \ldots, t_{M-1}\}$. \\

\noindent The static hedge portfolio payoff vector (hidden layer output) and output of the neural network at each simulation path $\omega_j$ are $\phi \big(S_{t}(\omega_j),\boldsymbol{b} \big) \in \mathbb{R}^{p}$ and $\tilde{G}^{\beta_{t}} \big(S_t(\omega_j)\big) = \phi^{\intercal} \big(S_{t}(\omega_j), \boldsymbol{b} \big)W$ respectively. In contrast, the target option portfolio value is $V_{t}\big(S_t(\omega_j)\big)$. Further, we define, 

\begin{itemize}
\item Hidden Layer Output (static hedge portfolio constituent options' payoff): \\
$X_{t}(\boldsymbol{b}) := \Big( \phi \big(S_{t}(\omega_j), \boldsymbol{b} \big) \Big)^{\intercal}_{j=1, .., N} \in \mathbb{R}^{N \times p}$, \\

\item Therefore, the static hedge portfolio value for the simulated paths is given as: \\
$X_{t}(\boldsymbol{b})W \in \mathbb{R}^{N}$ and, \\
\item Target Portfolio Value: \\
$Y_{t} = \Big( V_{t} \big(S_{t}(\omega_j) \big) \Big)^{\intercal}_{j=1, 2, ..., N} \in \mathbb{R}^{N}$.  \\
\end{itemize}

\noindent Therefore, by training the neural network to minimise $||Y_{t} - X_{t}(\boldsymbol{b}) W||^{2}$, with respect to both $W$ and $\boldsymbol{b}$, we can determine the strike prices and weights associated with the $p$ constituent options. In simpler terms, this process reveals the composition of the static hedge portfolio. This portfolio has to be set up at time $t_{m-1}$ and will act as the static hedge portfolio for the target Bermudan option from time $t_{m-1}$ to $t_m$. The static hedge portfolio can be valued at any time in the interval $(t_{m-1}, t_{m}]$ as a weighted sum of each constituent option value, priced by the Black-Scholes pricing model \cite{black1973pricing}. The value of the static hedge portfolio at time $t_{m-1}$ is the continuation value of the target Bermudan option.  \\

\subsection{Parameter Initialisation}
\label{Parameter Initialisation}
This section illustrates the initialisation of the model parameters $W$ and $\boldsymbol{b}$ for each neural network learnt at time horizons $t \in \{t_1, t_2, \ldots t_{M}\}$. \\

\noindent The parameter vector $\boldsymbol{b}$ is first chosen as equidistant $p$ strikes between out-of-the-money (OTM) and in-the-money (ITM) strikes, where moneyness is defined as the ratio of strike and spot at time $t_0$. This paper chooses strikes between $90\%$ and $110\%$ moneyness. But, this is subject to the market's liquidity.  It is noted that call and put option strikes are chosen separately between OTM and ITM strikes and considered together as initial strike vector $\boldsymbol{b}(0)$ to represent corresponding biases of call and put option hidden nodes. \\

At an exercise time $t$, by fixing the $p$ initialised strikes, $W$ which minimises $||Y_{t} - X_{t}(\boldsymbol{b}) W||^{2}_{\boldsymbol{b}=\boldsymbol{b}(0)}$ is used as the initial weight vector $W(0)$ of the neural network. This is achieved by linear regression (least-squares method without a constant term) between $\phi(S_{t}, \boldsymbol{b})$ as regressor variables and $V_{t}(S_t)$ as the response variable, and the regression coefficients are considered as optimal initialisation for $W$. \\

\subsection{Optimisation Methodology}
\label{Optimisation Methodology}

\noindent The neural network parameters are learnt by the back-propagation algorithm (refer \citeauthor{bishop1995neural} (\citeyear{bishop1995neural})) with the proposed optimisation methodology discussed in the next paragraph to iteratively reach the local/global minima. The neural network is trained using multiple epoch runs\footnote{one epoch is when the training data is used once}. Each epoch has multiple batches of fixed-size data randomly selected with replacement from the simulated underlying paths of training data (of size $N$). Therefore, the total number of iterative steps to learn the model parameters equals the product of the number of batches within each epoch and the number of epochs. In our proposed methodology, starting with initialised values of $W$ and $\boldsymbol{b}$, we move iteratively towards the optimal solution. \\

\noindent Let us define the loss function at any exercise time $t$ as, \\

 $L(t; W, \boldsymbol{b})$ $=$ $\frac{1}{2} \cdot \frac{1}{N} \cdot ||Y_t - X_{t}(\boldsymbol{b}) W||^{2}$, \\ 

\noindent and, for each path $\omega_j$, the pathwise loss function is, \\

 $L(t, \omega_j; W, \boldsymbol{b})$ $= \frac{1}{2} \cdot \Big[V_t\big(S_t(\omega_j) \big) - \big(\phi\big( S_t(\omega_j), \boldsymbol{b} \big) \big)^{\intercal} W \Big]^{2}$.\\


Let $W(l) = [w_1(l), w_2(l), ... w_p(l)]^{\intercal}$ and $\boldsymbol{b} (l) = [b_1(l), b_2(l), ...., b_p(l)]^{\intercal}$ be the static hedge portfolio weights and strikes vector respectively after the $l^{th}$ iteration step.

At each iteration, the loss function $L(t; W, \boldsymbol{b})$ is minimised by shifting $\boldsymbol{b},$ and $W$ using the Adaptive Moment Estimation (Adam) optimisation introduced by \citeauthor{kingma2014adam} (\citeyear{kingma2014adam}) to obtain the next strike vector $\boldsymbol{b} (l+1),$ and weights $W(l+1)$. This is a first-order method in both $\boldsymbol{b},$ and $W.$ We define a modified loss function that is, given $\boldsymbol{b}$ optimal in $W$. More precisely, 
\begin{align}
L^{*}\big(t, \omega_j; \boldsymbol{b}\big) := L\big(t, \omega_j; W^{*}, \boldsymbol{b}\big),
\end{align}
\noindent such that,
\begin{align}
 W^{*} &= \underset{W} {\mathrm{argmin}} \ \ L\big(t, \omega_j; W, \boldsymbol{b} \big).
\end{align}\label{opt-iter}

\noindent When we want to minimize  $L^{*}\big(t, \omega_j; \boldsymbol{b}\big),$ using Adam, we only shift the strikes $\boldsymbol{b}$ along its gradient with respect to $L^*.$ With $\boldsymbol{b}$ fixed, $L(\cdot)$ being linear in $W,$ $W^*$ in Equation \ref{opt-iter} can be obtained using the ordinary least squares regression. The iterations continue till the pre-determined maximum number of epoch runs are complete or if the stopping criteria\footnote{If the absolute mean error difference between two consecutive steps, i.e., absolute mean of $\big[L^*\big(t, \omega; \boldsymbol{b}(l+1)\big) - L^*\big(t, \omega; \boldsymbol{b}(l)\big)\big]$ is less than $10^{-8}$ for ten times in a row.} is reached. \\


\begin{enumerate}

\item Adam Optimisation for parameter $\boldsymbol{b}$: 

\noindent Let the partial derivative of the loss function with respect to the strikes vector $\boldsymbol{b}$ for the simulated path $\omega_j$ be, \\

\begin{align}
\frac{\partial L^*(t, \omega_j;  \boldsymbol{b})}{\partial \boldsymbol{b}} = \Big(\frac{ \partial L^*(t, \omega_j;  \boldsymbol{b}) }{\partial b_i}\Big)^{\intercal}_{i=1,2,\ldots,p} \in \mathbb{R}^{p}
\end{align}

Let $\eta(l) \in \mathbb{R}^{p}$ and $\nu(l) \in \mathbb{R}^{p}$ be the first-moment and second-moment terms (after $l$ iterations) of the Adam optimization, such that, $\eta(0) = \nu(0)$ is a Zero vector in $\mathbb{R}^{p}$. \\

\begin{enumerate}

\item Gradient of the modified Loss function with respect to $\boldsymbol{b}$: \\

\noindent Let the optimal loss with respect to the strike vector $\boldsymbol{b}$ (given a fixed $W$) for the simulated path $\omega_j$ be $L^{*}\big(t, \omega_j; \boldsymbol{b}\big)$. \\

\noindent The path-wise gradient of the modified loss function $L^*$ with respect to each strike $b_i$ after $l$ iterations is defined as,
\begin{align} \label{bp_grad}
\frac{\partial L^{*}(t, \omega_j; \boldsymbol{b})}{\partial b_i} \Big|_{\boldsymbol{b}=\boldsymbol{b}(l)} := \lim_{h \to 0} \frac{L^{*}\big(t, \omega_j; \boldsymbol{b}(l_{i+})\big) - L^{*}\big(t, \omega_j; \boldsymbol{b}(l_{i-})\big)}{2h}
\end{align}
\noindent where, 
\begin{align*}
\boldsymbol{b}(l_{i+}) &= [b_1(l), b_2(l),..., b_i(l) + h,..., b_p(l)]^{\intercal}, \\
\boldsymbol{b}(l_{i-}) &= [b_1(l), b_2(l),..., b_i(l) - h,..., b_p(l)]^{\intercal}.
\end{align*}

\noindent The gradient of the loss function $\frac{\partial L^{*}\big(t; \boldsymbol{b}\big)}{\partial \boldsymbol{b}}\Big|_{\boldsymbol{b}=\boldsymbol{b}(l)}$ for a batch $\mathcal{B}$ to be used in the iterative Adam optimisation algorithm to calculate the next  $\boldsymbol{b}(l+1)$ is given by,

\begin{align} \label{opt-gradient}
\frac{\partial L^{*}\big(t; \boldsymbol{b}\big)}{\partial \boldsymbol{b}}\Big|_{\boldsymbol{b}=\boldsymbol{b}(l)} = \sum_{j \in \mathcal{B}} \frac{\partial L^{*}(t, \omega_j; \boldsymbol{b})}{\partial \boldsymbol{b}}  \Big|_{\boldsymbol{b}=\boldsymbol{b}(l)}.
\end{align}

\item First and second-moment terms of Adam optimisation: \\
\begin{align*}
\eta(l+1) &= \beta_1 \cdot \eta(l) + (1-\beta_1) \cdot \Bigg( \frac{\partial L^{*}\big(t; \boldsymbol{b}\big)}{\partial \boldsymbol{b}}\Big|_{\boldsymbol{b}=\boldsymbol{b}(l)}  \Bigg),  \\
\nu(l+1) &= \beta_2 \cdot \nu(l) + (1-\beta_2) \cdot \Bigg( \frac{\partial L^{*}\big(t; \boldsymbol{b}\big)}{\partial \boldsymbol{b}}\Big|_{\boldsymbol{b}=\boldsymbol{b}(l)}  \Bigg)^{2},  \\
\eta_{cap} &= \frac{\eta(l+1)}{1 - \beta_1^{(l+1)}}, \\
\nu_{cap} &= \frac{\nu(l+1)}{1 - \beta_2^{(l+1)}}, 
\end{align*}

\noindent where, $\beta_1 = 0.9$ and $\beta_2=0.99$. \\

\item Iterative algorithm for the next optimal strike vector: 
\begin{align}\label{adam}
\boldsymbol{b}(l +1) &=  \boldsymbol{b}(l) - \eta_{cap} \cdot lr \cdot \Bigg(\frac{1}{\sqrt{\nu_{cap}}+ \epsilon} \Bigg),
\end{align}
\noindent where, the learning rate $lr = 0.001$ and $\epsilon = 10^{-8}$. Each strike is floored to a minimum value of $10^{-8}$ after every update. \\

\end{enumerate}

\item Linear Regression for parameter $W$: 
\begin{align}\label{lr}
W^*(l+1) &= \underset{W}{\mathrm{argmin}} \ \ L\big(t; W, \boldsymbol{b}(l+1)\big), \nonumber \\
 &= \Big(X_{t}^{\intercal}\big(\boldsymbol{b}(l+1)\big) \ X_{t}\big(\boldsymbol{b}(l+1)\big) \Big)^{-1} \ X_{t}^{\intercal}\big(\boldsymbol{b}(l+1)\big) \ Y_t.
\end{align}
\end{enumerate}

%
%
%


\begin{algorithm}[H]
\caption{Optimal Regress-later with Neural Networks (OPT-RLNN)}\label{alg-opt-rlnn}
\begin{algorithmic}[1]
\State Setup the target portfolio information (with strike $K$ and exercise points $\{t_0, t_1, t_2, \ldots, t_M\}$) and time-zero market data ($S_0$, $r$, $\sigma$)
\State Generate $S_{t_m}\left(\omega_j\right)$  for paths $j=1,\ldots,N, \, m = 0,\ldots,M$ 
\State $V_{t_M} \leftarrow h \left(S_{t_M}\right)$  evaluate option value for each path at $t_M$
\For{$m=M\ldots,1$}
	\State Initialize $\beta_{t_m}$ as per the proposed parameter initialisation (i.e. static hedge portfolio weights and constituent option strikes) in Section \ref{Parameter Initialisation}
       \State  $\beta_{t_m} \leftarrow \underset{\beta} {\mathrm{argmin}} \left(\frac{1}{N} \sum_{j = 1}^{N} \frac{1}{2} \left(V_{t_m}(\mathbf{S}_{t_m}(\omega_j))-G^{\beta}\left(\mathbf{S}_{t_m}(\omega_j)\right)\right)^2\right) = \underset{W, \boldsymbol{b}} {\mathrm{argmin}} \ L(t_m; W, \boldsymbol{b})$; Fitting the network with the proposed optimisation technique in Section \ref{Optimisation Methodology}
\For{$j=1,\ldots,N$}
\State $Q_{t_{m-1}}\left(S_{t_{m-1}}(\omega_j)\right) \leftarrow \mathbb{E}\left[\tilde{G}^{\beta_{t_m}} (S_{t_m}) \mid S_{t_{m-1}} (\omega_j)\right]$, which is the continuation value evaluated using Black-Scholes pricing model      
       \If{$h(S_{t_{m-1}}(\omega_j)) > Q_{t_{m-1}}\left(S_{t_{m-1}}(\omega_j)\right)$}
    			\State$V_{t_{m-1}}(S_{t_{m-1}}(\omega_j)) \leftarrow h(S_{t_{m-1}}(\omega_j))$
  	  \Else
    			\State $V_{t_{m-1}}(S_{t_{m-1}}(\omega_j)) \leftarrow Q_{t_{m-1}}\left(S_{t_{m-1}}(\omega_j)\right)$
      \EndIf
      \EndFor
    \EndFor
 
\end{algorithmic}
\end{algorithm}

\section{Analysis and Results}

The objective of the analysis section is to provide evidence for the accuracy and fast convergence of the proposed neural network model for pricing Bermudan options and perform the benchmarking analysis to study its pricing performance and ancillary advantages like future exposure generation. In this analysis, Bermudan put options of 1-year maturity at three moneyness (ATM, ITM and OTM) levels with exercise points (also called risk horizons from CCR perspective) at every consecutive three months are considered. The underlying is simulated using Geometric Brownian Motion as defined in Equation \ref{GBMprocessOptimalRLNN} with 5000 paths for benchmarking exercise and 50,000 paths for error-convergence analysis at each time horizon of interest. All the performances are analysed using simulated underlying data (validation data) generated independently from the training data used for building or learning the model parameters. \\

\noindent The following is the summary of the analysis performed: \\

\begin{itemize}
\item \textbf{Error-Convergence Analysis}: We compare the error convergence of RLNN and RLNN-OPT (or the static hedge model) model price across several epochs to demonstrate the accelerated convergence of the proposed model. The price of the COS model (refer Appendix \ref{Appdx: COS Method for Bermudan Option Pricing}) is considered the reference or accurate price to measure the error. \\
\item \textbf{Performance and Benchmarking Analysis}: In this section, we compare the PV distributions and exposures (under risk-neutral and real-world measure) generated by the static hedge model against the Longstaff Schwartz method (refer Appendix \ref{Appdx: Analysis of Interpolation Schemas for Long-Schwartz Method}) with cubic polynomial basis function. The COS model is used as a reference model to validate the model's accuracy. The benchmarking exercise of the exposure profiles is carried out at both standard and non-standard finer risk horizons. The following analyses are performed:
\begin{enumerate}
\item PV distribution Benchmarking
\item Exposures Benchmarking - Risk Neutral Measure
\item Exposures Benchmarking - Real-World Measure
\item Exposures Benchmarking at Finer-Grid Risk Horizons \\
\end{enumerate}
\item \textbf{Interpolation Schemes for Longstaff-Schwartz Method}: Different interpolation methods are proposed and compared for the industry standard Longstaff-Schwartz method to understand the impact on the accuracy of exposures at non-standard risk horizons. 
\end{itemize}

\subsection{Error-Convergence Analysis}

\begin{figure}[!htb]
\begin{center}
\includegraphics[width=\textwidth, height=3in]{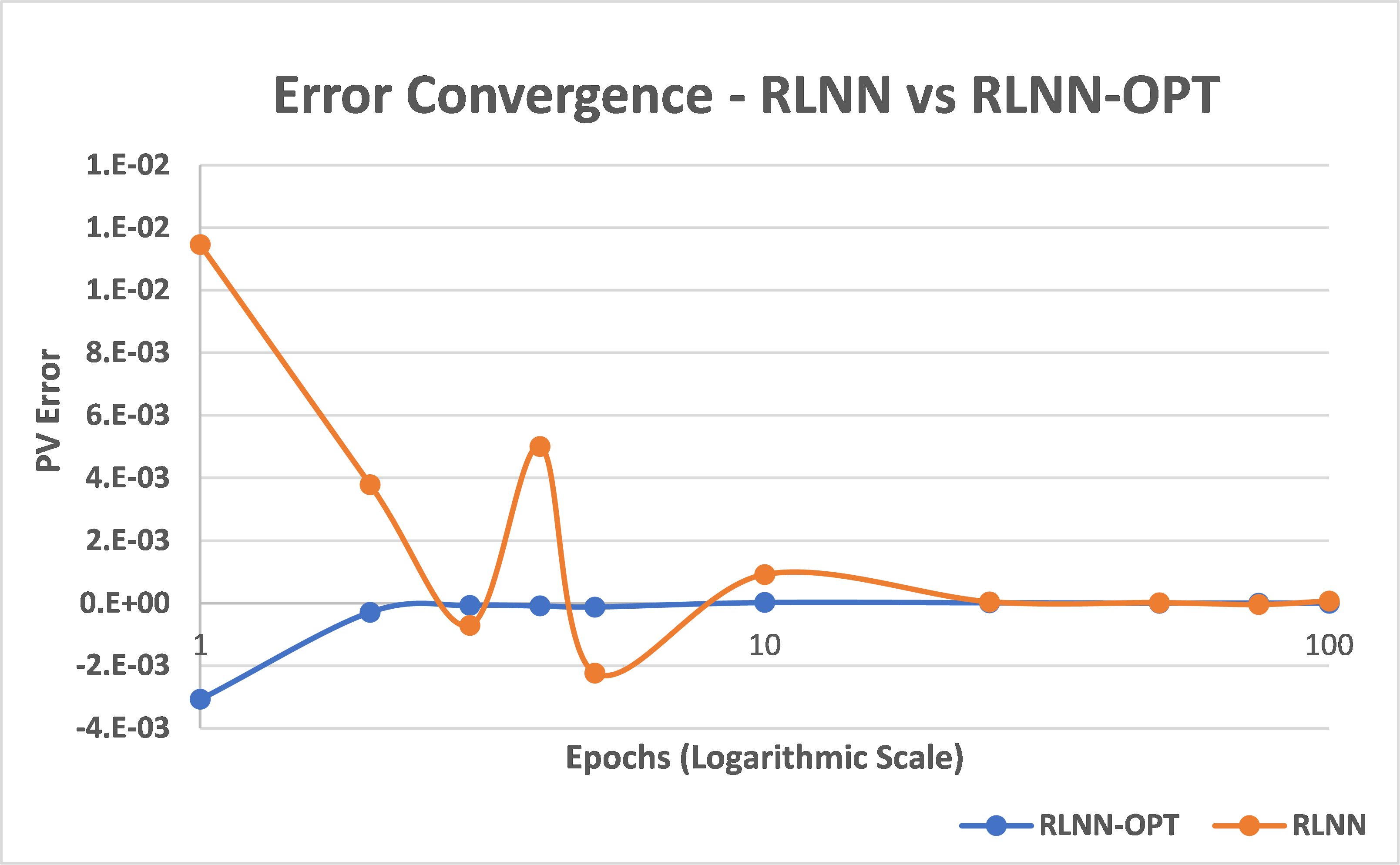}
\caption{\footnotesize The plot shows the model error convergence for RLNN and RLNN-OPT on pricing ATM Bermudan option. The x-axis corresponds to the number of epochs and employs a logarithmic scaling for spacing to have a magnified view at earlier epochs (though epoch values are exact). The y-axis corresponds to the PV error. The RLNN-OPT converges to the true price (i.e. with negligible error) in three epochs, whereas RLNN takes 25 epochs.}
 \label{Error_Convergence-ATM.png}
\end{center}
\end{figure}

\noindent In this section, we compare the error convergence of the proposed model ('RLNN-OPT' or 'Static Hedge Model') against the neural network model RLNN proposed by \citeauthor{lokeshwar2022explainable} (\citeyear{lokeshwar2022explainable}) for pricing Bermudan options across epochs. The RLNN model learns all the neural network parameters using the back-propagation algorithm by leveraging Adam optimisation. In the RLNN-OPT model (as detailed in Section \ref{Optimisation Methodology}), at each iteration, the static hedge portfolio strikes are shifted in the direction of the optimal gradient (defined in Equation \ref{opt-gradient}) that minimizes the loss function using the Adam optimiser. Subsequently, the portfolio weights are learned by regressing the target option value against payoffs of the constituent options of the static hedge portfolio. In this analysis, the reference price is obtained using the COS method. The error of a model is defined as the difference between the model's price and the COS model's price. In Figure \ref{Error_Convergence-ATM.png}, we present the errors of RLNN and RLNN-OPT models for pricing the ATM Bermudan option across epochs. The x-axis employs a logarithmic scaling for spacing while displaying the exact epoch values. We observe that RLNN-OPT converges to true price in three epochs, whereas RLNN takes 25 epochs to converge to true value, thereby highlighting the faster convergence capability of RLNN-OPT. Similar accelerated convergence with RLNN-OPT was observed in both ITM and OTM cases of the target option. 

\subsection{Benchmarking and Performance Analysis}
\subsubsection{PV distribution Benchmarking}

\begin{figure}[!htb]
\begin{center}
\includegraphics[width=0.9\textwidth, height=5.8in]{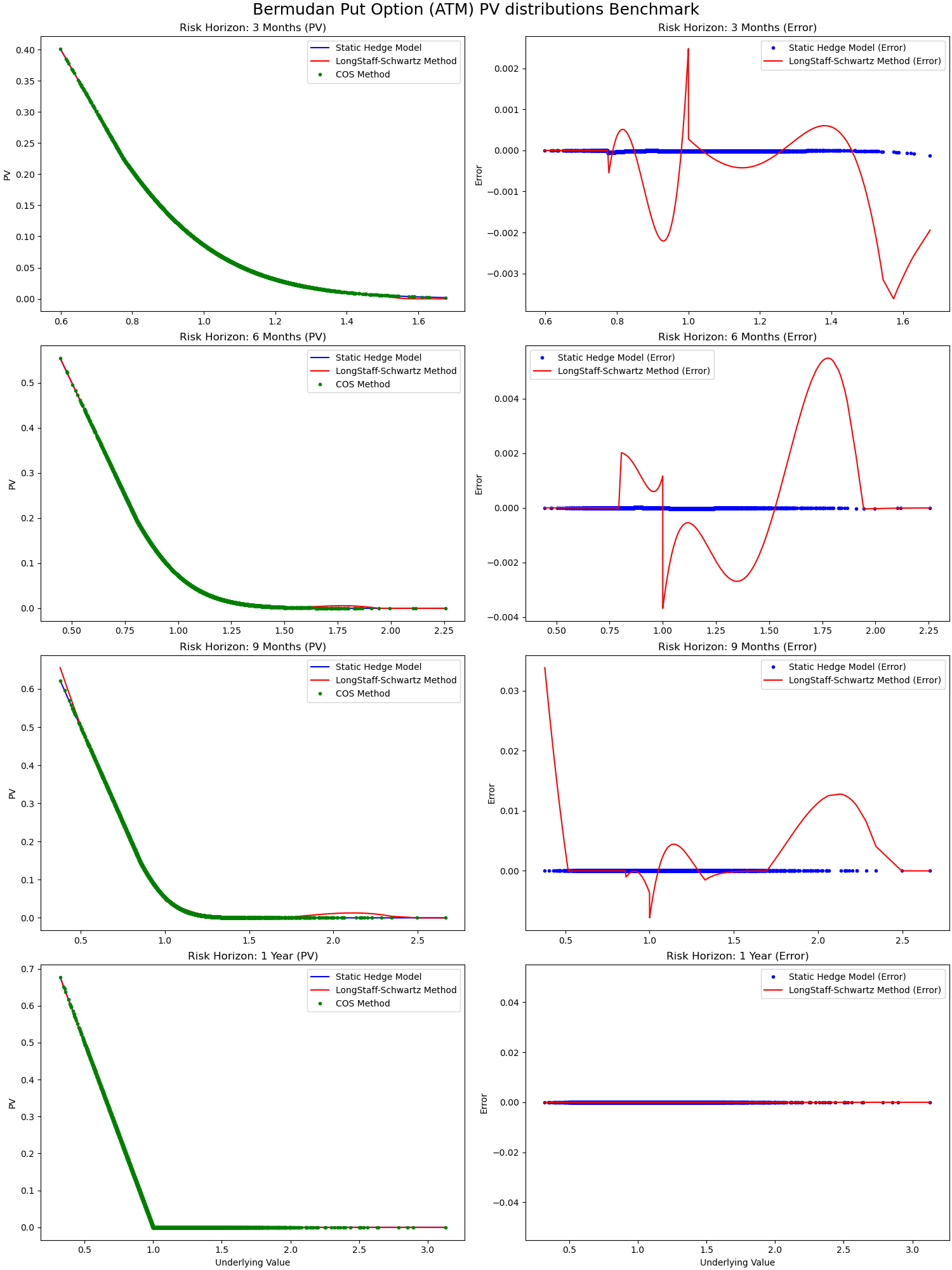}
\caption{\footnotesize The plot shows the PV distribution of ATM Bermudan Option across all future exercise time points obtained by the three pricing models: Static hedge model, the Longstaff-Schwartz model and the COS model on the left column. The right column shows the corresponding PV error for the static hedge model and the Longstaff-Schwartz model with respect to the COS model. The x-axis of all the subplots correspond to underlying levels, whereas the y-axis of the subplots on the left column correspond to PV and the right column to PV error.}
 \label{A2_bermudan_ATM_pv_dist_unadj.png}
\end{center}
\end{figure}

In this section, the comparison of the Bermudan option's PV distribution by the static hedge model against the Longstaff Schwartz model at each exercise time horizon is performed. The Longstaff-Schwartz method employed a cubic polynomial function as the chosen basis function. The comparison is assessed with the COS model as the reference model. \\

We performed this assessment at ATM ($100\%$), ITM ($110\%$) and OTM ($90\%$) moneyness levels (measured as a ratio of strike to spot for a put option) of the target option. The performance results for ATM options can be found in Figure \ref{A2_bermudan_ATM_pv_dist_unadj.png}; similar observations were seen with ITM and OTM options. The left column of the plot shows the distribution of PV of the three models of interest at different exercise points. The right column shows the error (defined as the difference of PV with respect to the COS model price) of the static hedge model and Longstaff Schwartz method across simulated underlying levels for the corresponding exercise time point. The COS model PV distribution is generated by the option valuation using the COS model at all simulated paths and time horizons. Compared to the Longstaff method, the static hedge model is aligned with the COS model across all underlying levels. It can be noted that the static hedge model can replicate the COS model price of the target option better across the entire spectrum of the underlying when compared to the Long-Schwartz model. 

\subsubsection{Exposures Benchmarking - Risk Neutral Measure}

In this section, we compare the Expected Exposure (EE) and Potential Future Exposure (PFE) generated by the static hedge model and Longstaff-Schwartz method by benchmarking against exposure profiles by the COS method. It is to be noted that both the scenario generation of the underlying and pricing are performed under risk-neutral measure. Firstly, the exposure distribution is computed, and then the EE and PFE (at $99\%$ confidence level) are calculated at each risk horizon. It is to be noted that the PV distribution seen in the previous section is different from the Exposure distribution. The PV distribution is the distribution of option value at a specific time horizon conditioned on no exercise before. When exposure distribution is calculated, if the option has been exercised before the current time horizon in a simulated path, the exposure at the current and future time in the respective path is zero. \\ 

\noindent Therefore, Exposure $EXP[t_m; S_{t_m}(\omega_j)]$ at a certain path $\omega_j$ and time horizon $t_m$ is, 

\[
    EXP[t_m; S_{t_m}(\omega_j)]= \begin{cases}
         V_{t_m}\big(S_{t_m}(\omega_j)\big),& \text{if } \tau(\omega_j) \geq t_m \\
        0,              & \text{otherwise}
    \end{cases}
\]

and, the stopping time $\tau(\omega_j)$ for the path $\omega_j$ is, 
\begin{align*}
\tau(\omega_j) = \min \{t \in [t_1, t_2, \ldots, t_M=T]: h\big(S_{t}(\omega_j)\big) > Q_{t}\big(S_{t}(\omega_j)\big) \}
\end{align*}

\begin{figure}[htbp!]
\begin{center}
\includegraphics[width=\textwidth, height=3in]{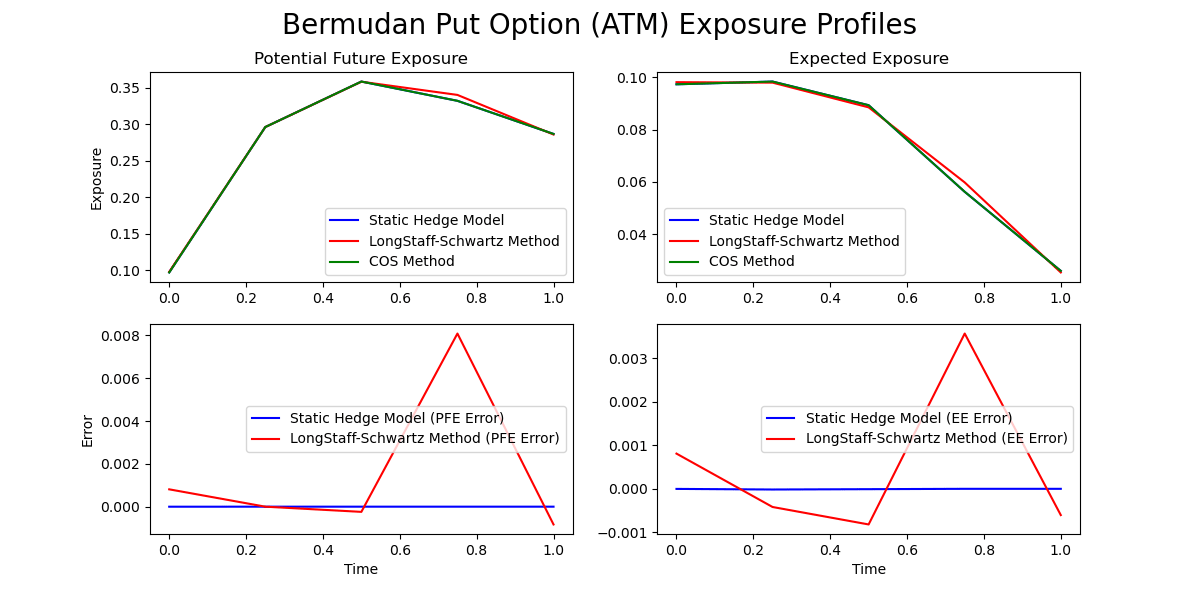}
\caption{\footnotesize The figure shows the exposure profiles of ATM Bermudan Option under risk-neutral measure at the exercise time points. In the first row, the left subplot corresponds to Potential Future Exposure, and the right plot corresponds to Expected Exposure. The exposures are generated by the Static hedge model, Longstaff-Schwartz model and COS model. The second row highlights the corresponding errors of the static hedge model and Longstaff-Schwartz model with respect to the COS model. The y-axis of the subplots in the first row corresponds to the exposure value. The y-axis of the subplots in the second row corresponds to the error value. The x-axis of all plots corresponds to time (in years).} \label{A2_bermudan_ATM_exposure_adj_benchmark.png}
\end{center}
\end{figure}

\noindent We present the exposure profile benchmarking results for ATM Bermudan put option in Figure \ref{A2_bermudan_ATM_exposure_adj_benchmark.png} and in Appendix \ref{Apdx: Exposures Benchmarking - Risk Neutral Measure} for ITM and OTM options. Overall, we find that the exposure profiles of the static hedge model align with the COS method. Further, the static hedge model is consistently better than the Longstaff method in all the cases considered.

\subsubsection{Exposures Benchmarking - Real World Measure}

\begin{figure}[htbp!]
\begin{center}
\includegraphics[width=\textwidth, height=3in]{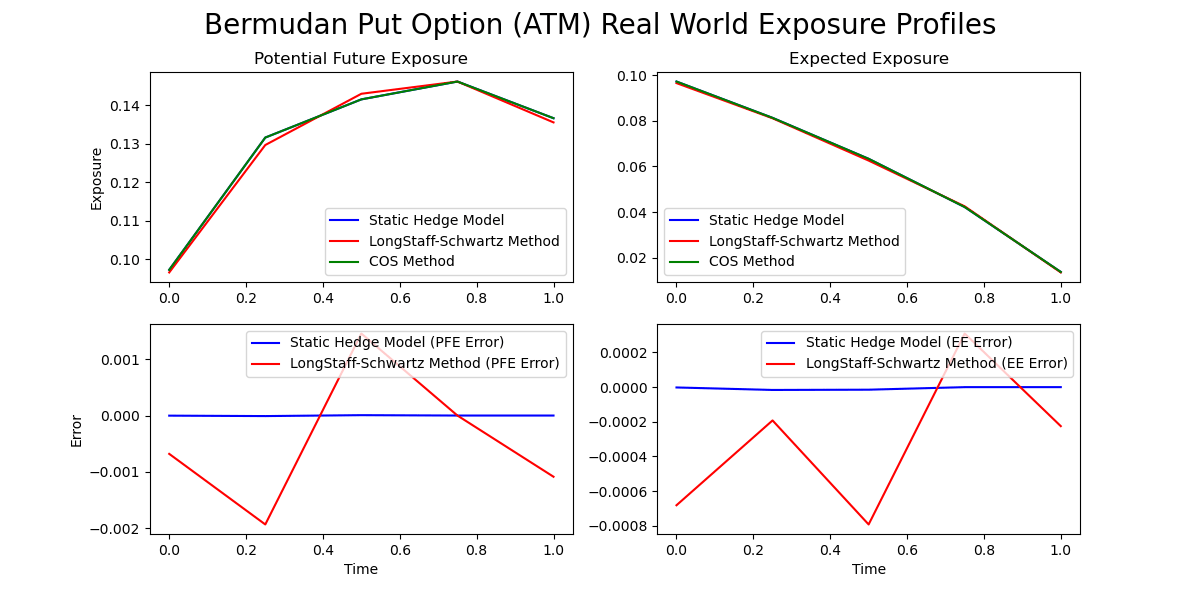}
\caption{\footnotesize The figure shows the exposure profiles of ATM Bermudan Option under real-world scenario 1 ($S_0 =1 , \sigma_{real}=0.1, \mu=0.07$)  at the exercise time points. In the first row, the left subplot corresponds to Potential Future Exposure, and the right plot corresponds to Expected Exposure. The exposures are generated by the Static hedge model, Longstaff-Schwartz model and COS model. The second row highlights the corresponding errors of the static hedge model and the Longstaff-Schwartz model with respect to the COS model. The y-axis of the subplots in the first row corresponds to the exposure value. The y-axis of the subplots in the second row corresponds to the error value. The x-axis of all plots corresponds to time (in years).} \label{A2_bermudan_ATM_exposure_adj_benchmark_real_scenario1.png}
\end{center}
\end{figure}

\noindent Few banks calculate the CCR risk statistics like PFE under real-world measure while other banks stick with risk-neutral measure. In this section, the capability of the proposed model to generate risk statistics under real-world measure is discussed. There are four stressed time zero market data scenarios, which are considered real-world market data scenarios to benchmark the exposure profiles. The four real-world time-zero market scenarios are:

\begin{itemize}
\item Scenario 1: $S_0 =1 , \sigma_{real}=0.1, \mu=0.07$
\item Scenario 2: $S_0 =1 , \sigma_{real}=0.3, \mu=0.1$
\item Scenario 3: $S_0 =1 , \sigma_{real}=0.5, \mu=0.15$
\item Scenario 4: $S_0 =1 , \sigma_{real}=0.5, \mu=0.01$

\end{itemize}

\noindent The GBM process generates the future underlying distribution with assumed underlying returns $\mu$ and volatility $\sigma_{real}$ corresponding to the four market scenarios. The three models of interest are used to generate real-world exposures. It is noted that the scenario generation of the underlying is performed under real-world measure, whereas, pricing is performed using parameters of the models learnt under risk-neutral measure. In other words, the underlying simulated paths are generated under real-world measure and pricing is performed under risk-neutral measure at each simulated path. From Figure \ref{A2_bermudan_ATM_exposure_adj_benchmark_real_scenario1.png} (Scenario 1 of ATM case) and Appendix \ref{Appdx: Exposures Benchmarking - Real World Measure} (for Scenarios 2, 3 \& 4 of ATM case), we observe that both the exposure profiles (EE \& PFE) of static hedge model closely align across all risk horizons with COS method. The static hedge model exhibits consistently better behaviour than the Longstaff-Schwartz model. A similar performance was observed for the ITM and the OTM case.

\subsection{Exposures Benchmarking at Finer-Grid Risk Horizons}

\begin{figure}[!htb]
\begin{center}
\includegraphics[width=\textwidth, height=3in]{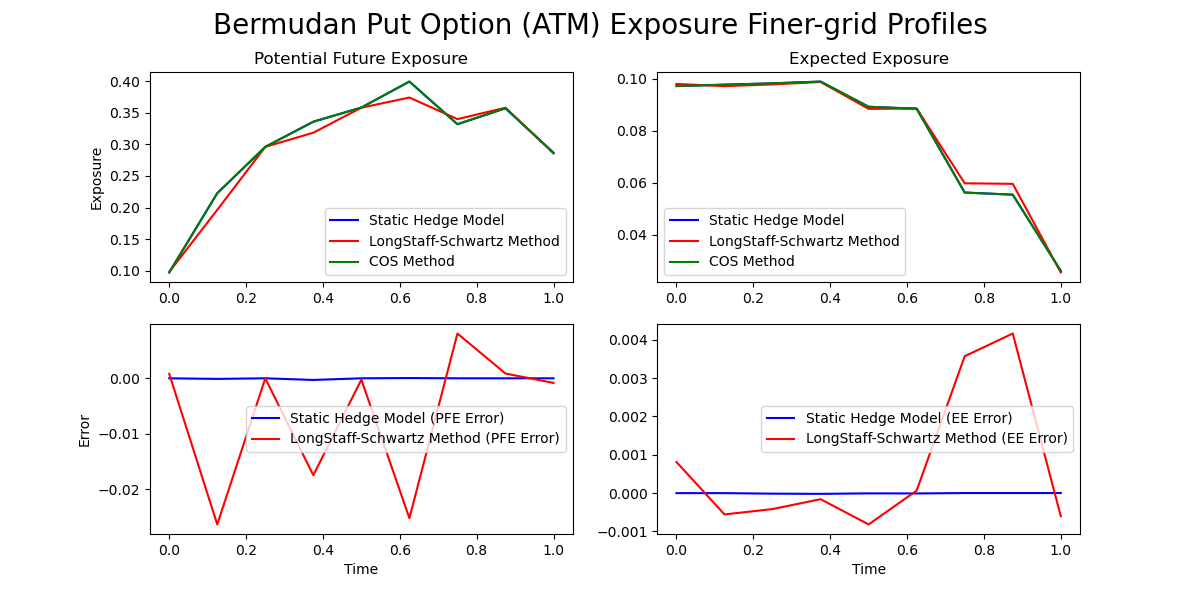}
\caption{\footnotesize The figure shows the exposure profiles of ATM Bermudan Option under risk-neutral measures at both exercise and finer grid horizons. In the first row, the left subplot corresponds to Potential Future Exposure, and the right plot corresponds to Expected Exposure. The exposures are generated by the Static hedge model, Longstaff-Schwartz model and COS model. The second row highlights the corresponding errors of the static hedge model and Longstaff-Schwartz model with respect to the COS model. The y-axis of the subplots in the first row corresponds to the exposure value. The y-axis of the subplots in the second row corresponds to the error value. The x-axis of all plots corresponds to time (in years).} \label{A2_bermudan_ATM_exposure_adj_finegrid_benchmark.png}
\end{center}
\end{figure}

In many applications, exposures at non-standard or non-exercise points are required. For example, it is generally preferred to use Front-Office (FO) pricing libraries for risk system. The FO system and risk system could differ in pricing model, market data and risk factors in the model. Enhancing the alignment between the front office and risk systems leads to stronger risk management practices. The FO pricing model parameters are designed to generate exposures at certain fixed time horizons. The horizons used in risk applications could be different, thereby, requiring generating exposures at non-standard points. This creates a necessity of generating exposures at intermediate points by interpolation or other approximations. Further, as discussed earlier, the calculation of margined exposures under cash collateral requires evaluation of PnL distribution over the margin period of risk. This mandates generation of PV distribution at additional points to calculate PnL. The proposed model provides an efficient solution to generate exposure at non-standard risk horizons. We simulate the underlying prices at the intermediate points and get the static hedge model price (by valuing the static hedge portfolio by the Black-Scholes pricing model) and COS model price at each intermediate point. For the Longstaff-Schwartz method, the value at the intermediate point is obtained through linear interpolation between the option value at the preceding and succeeding time horizons for each simulated path. The risk profiles inclusive of this intermediate time horizon are compared in Figure \ref{A2_bermudan_ATM_exposure_adj_finegrid_benchmark.png} (ATM case) and Appendix \ref{Appdx: Exposures at Finer-Grid} (ITM and OTM case). We observe a close alignment of the static hedge model with the COS model. There is some misalignment for the Longstaff Schwartz model, which is expected as the option values are directly interpolated, and the stock price variations between two consecutive exercise points could impact the accuracy. The following section studies different interpolation schemes for the Longstaff Schwarz method.

\subsubsection{Longstaff-Schwartz Method: Interpolation Schemas}

Longstaff-Schwartz method is the widely adopted approach for pricing Bermudan options and for approximating future exposure distributions. In this section, we study the performance of different interpolations to generate exposures at intermediate points between two exercise dates under Longstaff-Schwartz modelling framework. The objective is not to benchmark with other pricing methods but to understand which interpolation best reflects Longstaff Schwartz exposure. Firstly, we take the true value at intermediate points by fitting the cubic polynomial as performed in exercise dates. In other words, with simulated underlying values, we fit a cubic polynomial between the discounted value at the next exercise point against intermediate underlying values. We call this True Fit in the results. 

\begin{figure}[!htb]
\begin{center}
\includegraphics[width=\textwidth, height=3in]{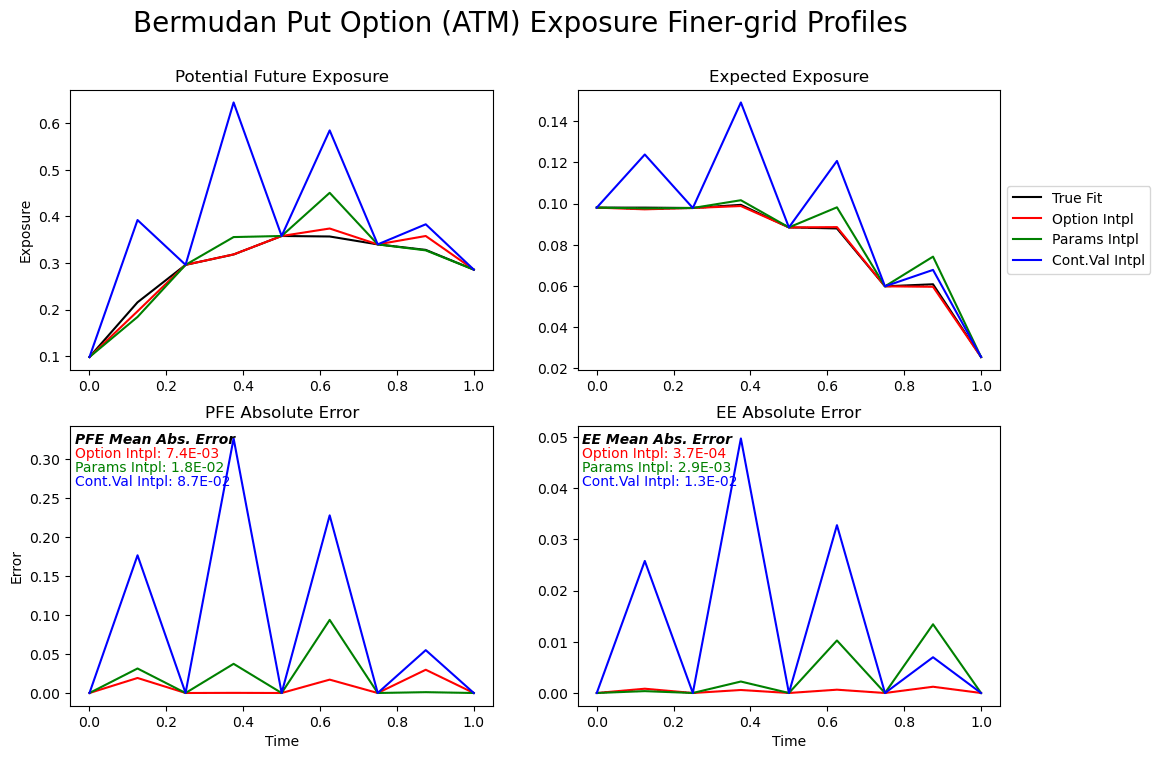}
\caption{\footnotesize The figure shows the exposure profiles of ATM Bermudan Option under risk-neutral measures at both exercise and finer grid horizons. In the first row, the left subplot corresponds to Potential Future Exposure, and the right plot corresponds to Expected Exposure. The exposures are generated by the Longstaff-Schwartz model using true fit (a cubic polynomial fit at finer horizons), option value interpolation (legend: Option Intpl), continuation value interpolation (legend: Cont.Val Intpl) and parameters interpolation (legend: Params Intpl). The second row highlights the corresponding interpolation errors with respect to the true model. The y-axis of the subplots in the first row corresponds to the exposure value. The y-axis of the subplots in the second row corresponds to the error value. The x-axis of all plots corresponds to time (in years).} \label{A2_bermudan_ATM_exposure_adj_finegrid_benchmark_longSchwarz.png}
\end{center}
\end{figure}

The three approaches we are looking at are: \\

\begin{itemize}
\item \textit{Option Value Interpolation}: In this method, we linearly interpolate the option values between two consecutive exercise dates to get option values at intermediate points. For any time $t \in (t_{m-1}, t_{m})$ for $m \in {1, 2, \ldots, N}$  and simulated path $\omega_j$, the intermediate price is 

\begin{align}
V_t\big(S_t(\omega_j)\big)  &=  V_{t_{m-1}}\big(S_{t_{m-1}}(\omega_j)\big) \nonumber \\ 
& + \Bigg[\frac{V_{t_{m}}\big(S_{t_{m}}(\omega_j)\big) - V_{t_{m-1}}\big(S_{t_{m-1}}(\omega_j)\big) }{t_m - t_{m-1}} \Bigg] \cdot (t - t_{m-1}).
\end{align}

It is to be noted that,  $V_{t_{0}}\big(S_{t_{0}}(\omega_j)\big) = V_{t_0}(S_{t_{0}})$ and $V_{t_{M}}\big(S_{t_{M}}(\omega_j)\big) = h\big(S_{t_M}(\omega_j)\big)$. \\

\item \textit{Continuation Value Interpolation}: In this method, we linearly interpolate the continuation values between exercise dates to determine the option value at intermediate points. Similar to the above scheme, the price at an intermediate point is, 

\begin{align}
 V_t\big(S_t(\omega_j)\big) &=  Q_{t_{m-1}}\big(S_{t_{m-1}}(\omega_j)\big) \nonumber \\ 
& + \Bigg[\frac{Q_{t_{m}}\big(S_{t_{m}}(\omega_j)\big) - Q_{t_{m-1}}\big(S_{t_{m-1}}(\omega_j)\big) }{t_m - t_{m-1}}\Bigg] \cdot (t - t_{m-1})
\end{align}

It is to be noted that,  $Q_{t_{0}}\big(S_{t_{0}}(\omega_j)\big) = V_{t_0}(S_{t_{0}})$ and $Q_{t_{M}}\big(S_{t_{M}}(\omega_j)\big) = h\big(S_{t_{M}}(\omega_j)\big)$. \\

\item \textit{Parameters Interpolation}: In this method, the cubic polynomial parameters are interpolated between two exercise days and evaluated using the underlying value at the intermediate time corresponding to each path. It is noted that the first intermediate point in the first Bermudan exercise period is the interpolation of the time-zero price and continuation value at the next time point. Similarly, the last finer grid point in the last Bermudan exercise period interpolates the continuation value at the previous exercise time and the intrinsic value at maturity. \\
\end{itemize}

\noindent Based on the comparison plots in Figure \ref{A2_bermudan_ATM_exposure_adj_finegrid_benchmark_longSchwarz.png} (ATM case) and Appendix \ref{Appdx: Analysis of Interpolation Schemas for Long-Schwartz Method} (ITM and OTM case), we observe that interpolation of option values tends to have consistently lower error compared to other interpolation schemes.

\section{Conclusion}
In this paper, we introduced an enhanced regress-later neural network framework tailored for the valuation and risk management of single-asset Bermudan options. The enhancement focuses on accelerating the convergence speed of the option price. This was achieved by developing RLNN-OPT neural network with an improved optimisation approach and parameter initialisation technique. The evidence for this enhancement is supported by conducting an error convergence analysis and comparing it with the neural network RLNN proposed by \citeauthor{lokeshwar2022explainable} (\citeyear{lokeshwar2022explainable}). For this analysis, the COS model is considered to be the reference model. \\

\noindent The proposed neural network possesses a meaningful interpretation, resembling a European call-and-put options portfolio. This interpretation helps to set up a self-replicating static hedge portfolio of the target Bermudan option. The bias terms associated with hidden nodes correspond to the strike prices of constituent options within the static hedge portfolio. The neural network weights connecting the hidden nodes to the outer layer correspond to the weightings of the constituent options they represent. \\

\noindent In the context of counterparty credit risk, anticipating future exposure distributions serves various pivotal purposes, including the computation of minimum capital requirements, credit limit monitoring, and more. Our proposed model generates accurate exposure distributions and profiles at no additional computational cost. We have additionally validated the accuracy of exposures generated by comparing our model against the Longstaff Schwartz method and benchmarking it against the COS method across all exercise dates. The need for exposure calculations at grid points beyond predefined risk horizons often arises. Our model accurately (relative to COS method) calculates exposures at non-standard risk horizons by leveraging the closed-form Black-Scholes pricing formula. Lastly, we comprehensively analysed various interpolation techniques within the Longstaff Schwartz framework to derive exposures at non-standard points.

\backmatter

\section*{Declarations}
The authors report no conflicts of interest. The authors alone are responsible for the content and writing of the paper.

\begin{appendices}





\section{Exposures Benchmarking - Risk Neutral Measure}
\label{Apdx: Exposures Benchmarking - Risk Neutral Measure}

\begin{figure}[!htb]
\begin{center}
\includegraphics[width=0.8\textwidth, height=2.2in]{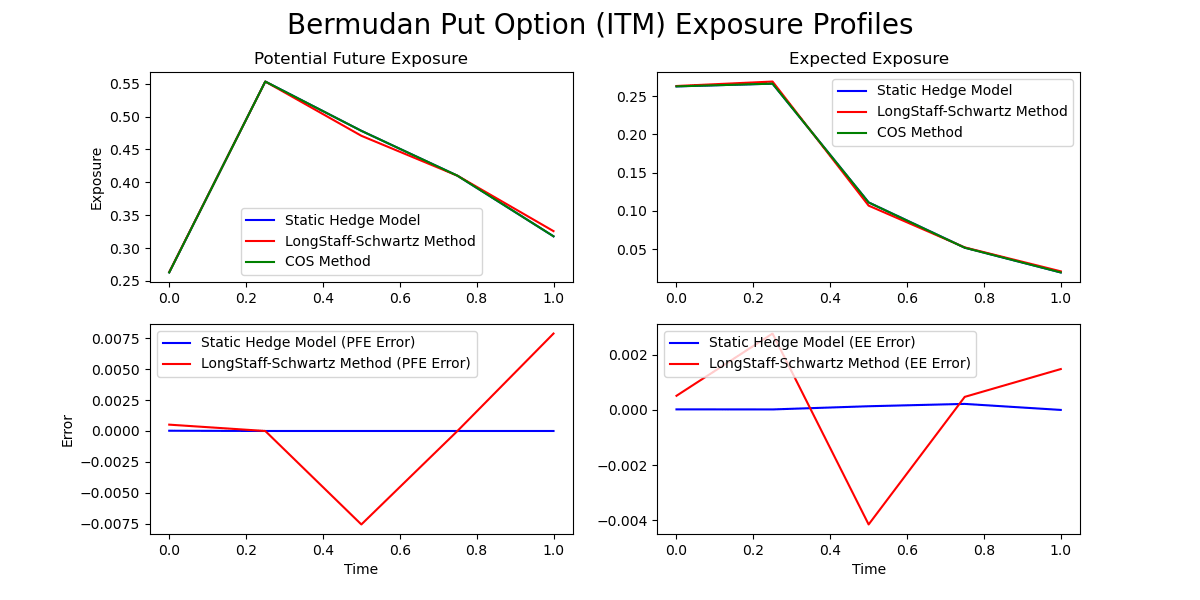}
\caption{\footnotesize The figure shows the exposures profiles of ITM Bermudan Option under risk-neutral measure at the exercise time points. In the first row, the left subplot corresponds to Potential Future Exposure and the right plot corresponds to the Expected Exposure. The exposures are generated by three models: Static hedge model, Longstaff-Schwartz model and COS model. The second row highlights the corresponding errors of static hedge model and Longstaff-Schwartz model with respect to COS model. The y-axis of the subplots in the first row corresponds to exposure value. The y-axis of the subplots in second row corresponds to error value. The x-axis of all plots correspond to time (in years)} \label{A2_bermudan_ITM_exposure_adj_benchmark.png}
\end{center}
\end{figure}

\begin{figure}[!htb]
\begin{center}
\includegraphics[width=0.8\textwidth, height=2.2in]{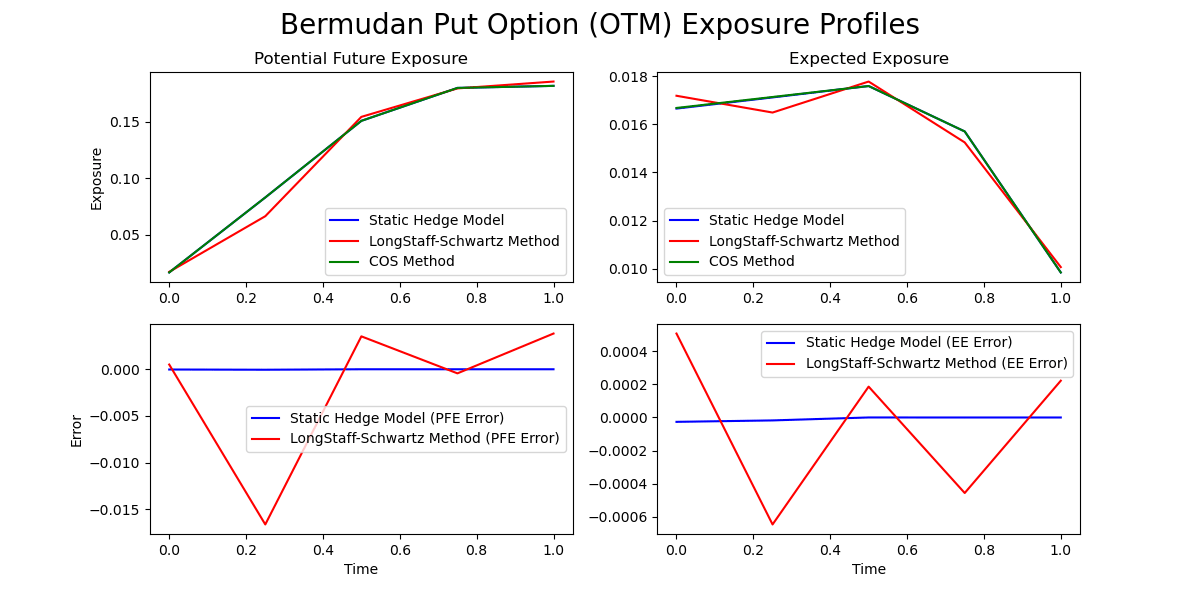}
\caption{\footnotesize The figure shows the exposures profiles of OTM Bermudan Option under risk-neutral measure at the exercise time points. In the first row, the left subplot corresponds to Potential Future Exposure and the right plot corresponds to the Expected Exposure. The exposures are generated by three models: Static hedge model, Longstaff-Schwartz model and COS model. The second row highlights the corresponding errors of static hedge model and Longstaff-Schwartz model with respect to COS model. The y-axis of the subplots in the first row corresponds to exposure value. The y-axis of the subplots in second row corresponds to error value. The x-axis of all plots correspond to time (in years)} \label{A2_bermudan_OTM_exposure_adj_benchmark.png}
\end{center}
\end{figure}

\section{Exposures Benchmarking - Real World Measure}
\label{Appdx: Exposures Benchmarking - Real World Measure}
\begin{figure}[!htb]
\begin{center}
\includegraphics[width=0.8\textwidth, height=2.2in]{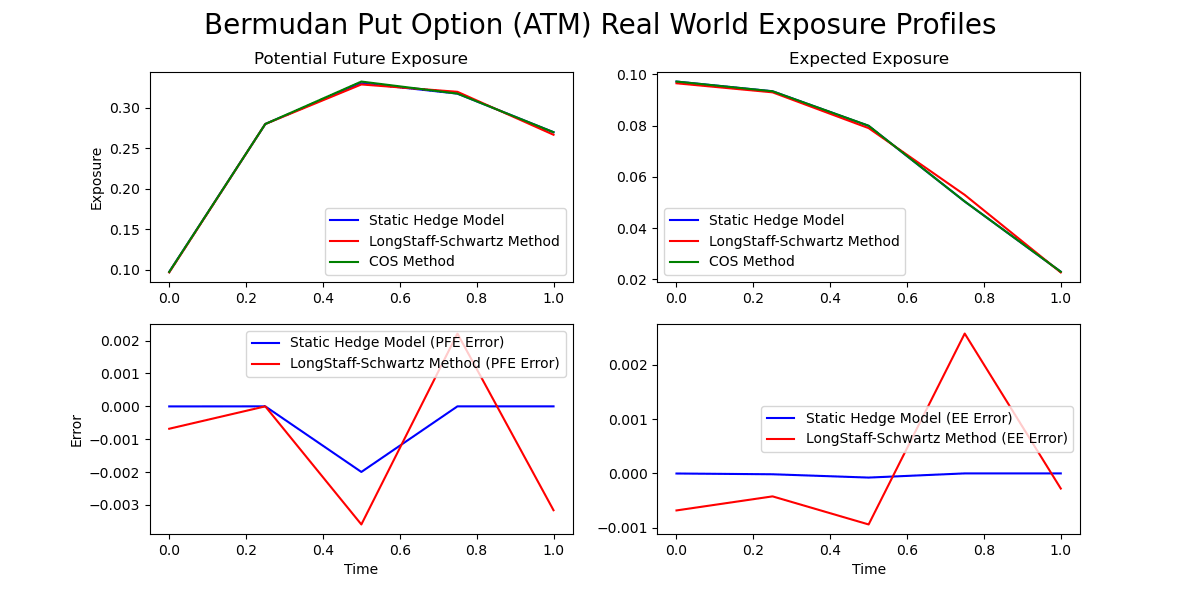}
\caption{\footnotesize The figure shows the exposures profiles of ATM Bermudan Option under real-world scenario 2 ($S_0 =1 , \sigma_{real}=0.3, \mu=0.1$)  at the exercise time points. In the first row, the left subplot corresponds to Potential Future Exposure and the right plot corresponds to the Expected Exposure. The exposures are generated by three models: Static hedge model, Longstaff-Schwartz model and COS model. The second row highlights the corresponding errors of static hedge model and Longstaff-Schwartz model with respect to COS model. The y-axis of the subplots in the first row corresponds to exposure value. The y-axis of the subplots in second row corresponds to error value. The x-axis of all plots correspond to time (in years).)} \label{A2_bermudan_ATM_exposure_adj_benchmark_real_scenario2.png}
\end{center}
\end{figure}

\begin{figure}[!htb]
\begin{center}
\includegraphics[width=0.8\textwidth, height=2.2in]{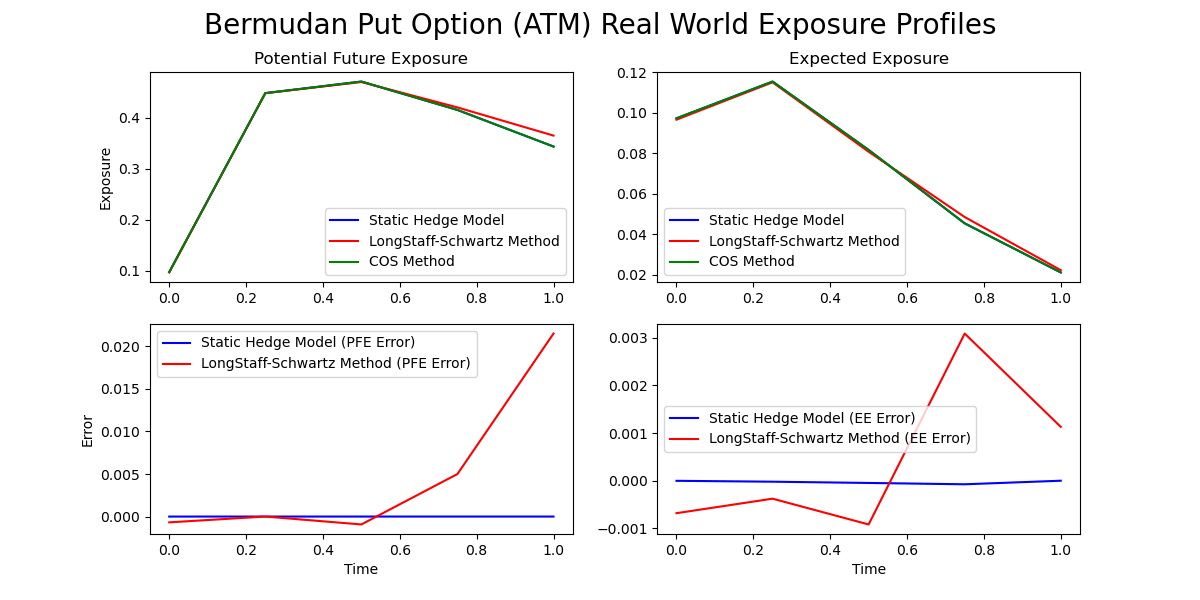}
\caption{\footnotesize The figure shows the exposures profiles of ATM Bermudan Option under real-world scenario 3 ($S_0 =1 , \sigma_{real}=0.5, \mu=0.15$)  at the exercise time points. In the first row, the left subplot corresponds to Potential Future Exposure and the right plot corresponds to the Expected Exposure. The exposures are generated by three models: Static hedge model, Longstaff-Schwartz model and COS model. The second row highlights the corresponding errors of static hedge model and Longstaff-Schwartz model with respect to COS model. The y-axis of the subplots in the first row corresponds to exposure value. The y-axis of the subplots in second row corresponds to error value. The x-axis of all plots correspond to time (in years).} \label{A2_bermudan_ATM_exposure_adj_benchmark_real_scenario3.png}
\end{center}
\end{figure}

\begin{figure}[!htb]
\begin{center}
\includegraphics[width=0.8\textwidth, height=2.2in]{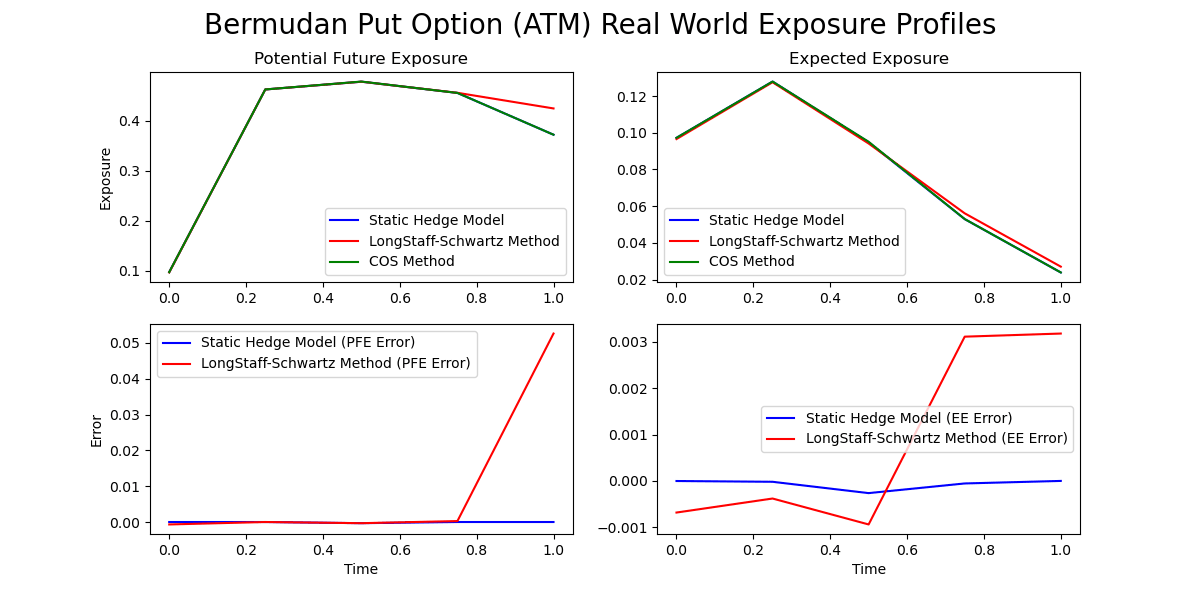}
\caption{\footnotesize The figure shows the exposures profiles of ATM Bermudan Option under real-world scenario 4 ($S_0 =1 , \sigma_{real}=0.5, \mu=0.01$)  at the exercise time points. In the first row, the left subplot corresponds to Potential Future Exposure and the right plot corresponds to the Expected Exposure. The exposures are generated by three models: Static hedge model, Longstaff-Schwartz model and COS model. The second row highlights the corresponding errors of static hedge model and Longstaff-Schwartz model with respect to COS model. The y-axis of the subplots in the first row corresponds to exposure value. The y-axis of the subplots in second row corresponds to error value. The x-axis of all plots correspond to time (in years).} \label{A2_bermudan_ATM_exposure_adj_benchmark_real_scenario4.png}
\end{center}
\end{figure}

\section{Exposures at Finer-Grid}
\label{Appdx: Exposures at Finer-Grid}

\begin{figure}[!htb]
\begin{center}
\includegraphics[width=0.8\textwidth, height=2.2in]{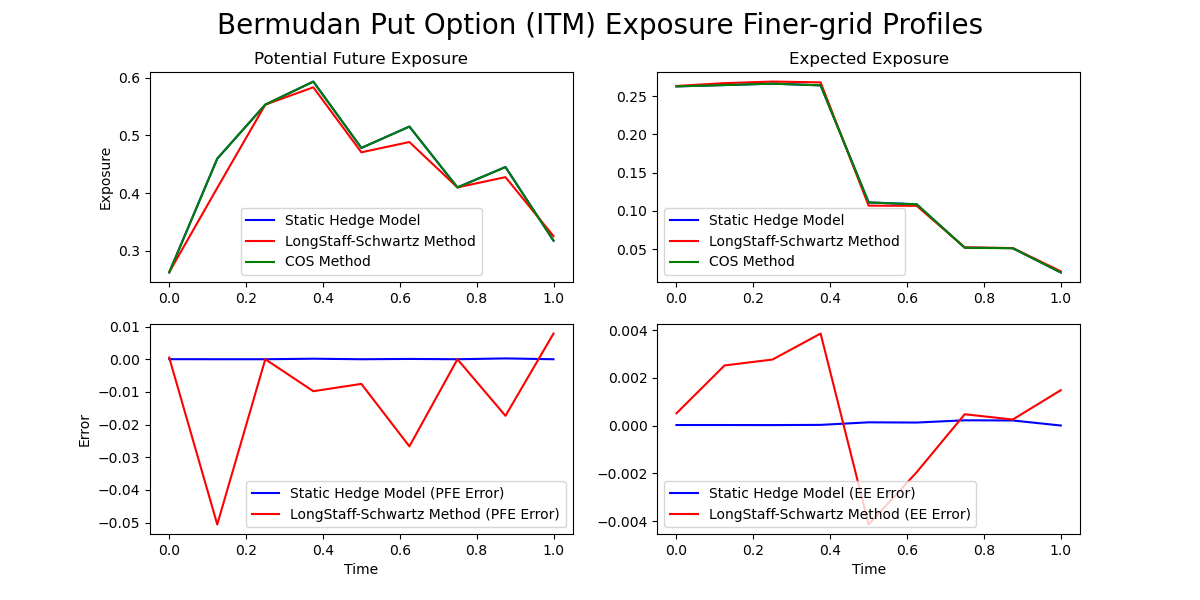}
\caption{\footnotesize The figure shows the exposure profiles of ITM Bermudan Option under risk-neutral measures at both exercise and finer grid horizons. In the first row, the left subplot corresponds to Potential Future Exposure, and the right plot corresponds to Expected Exposure. The exposures are generated by the Static hedge model, Longstaff-Schwartz model and COS model. The second row highlights the corresponding errors of the static hedge model and Longstaff-Schwartz model with respect to the COS model. The y-axis of the subplots in the first row corresponds to the exposure value. The y-axis of the subplots in the second row corresponds to the error value. The x-axis of all plots corresponds to time (in years).} \label{A2_bermudan_ITM_exposure_adj_finegrid_benchmark.png}
\end{center}
\end{figure}

\clearpage

\begin{figure}[!htb]
\begin{center}
\includegraphics[width=0.8\textwidth, height=2.2in]{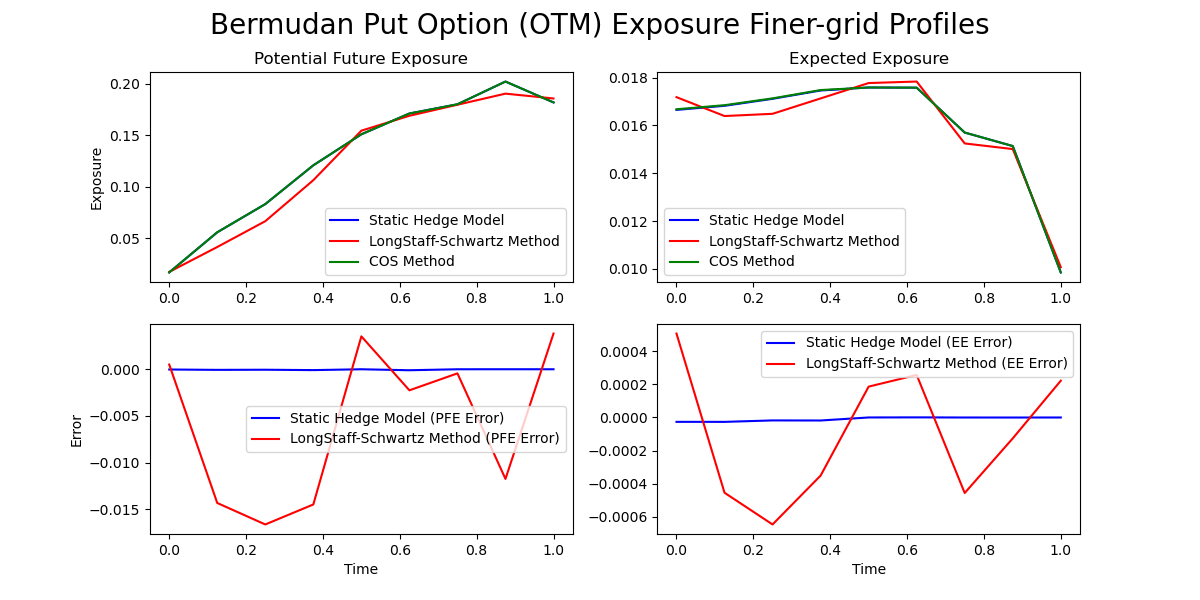}
\caption{\footnotesize The figure shows the exposure profiles of OTM Bermudan Option under risk-neutral measures at both exercise and finer grid horizons. In the first row, the left subplot corresponds to Potential Future Exposure, and the right plot corresponds to Expected Exposure. The exposures are generated by the Static hedge model, Longstaff-Schwartz model and COS model. The second row highlights the corresponding errors of the static hedge model and Longstaff-Schwartz model with respect to the COS model. The y-axis of the subplots in the first row corresponds to the exposure value. The y-axis of the subplots in the second row corresponds to the error value. The x-axis of all plots corresponds to time (in years).} \label{A2_bermudan_OTM_exposure_adj_finegrid_benchmark.png}
\end{center}
\end{figure}

\subsection{Analysis of Interpolation Schemas for Long-Schwartz Method}
\label{Appdx: Analysis of Interpolation Schemas for Long-Schwartz Method}

\begin{figure}[!htb]
\begin{center}
\includegraphics[width=0.8\textwidth, height=2.2in]{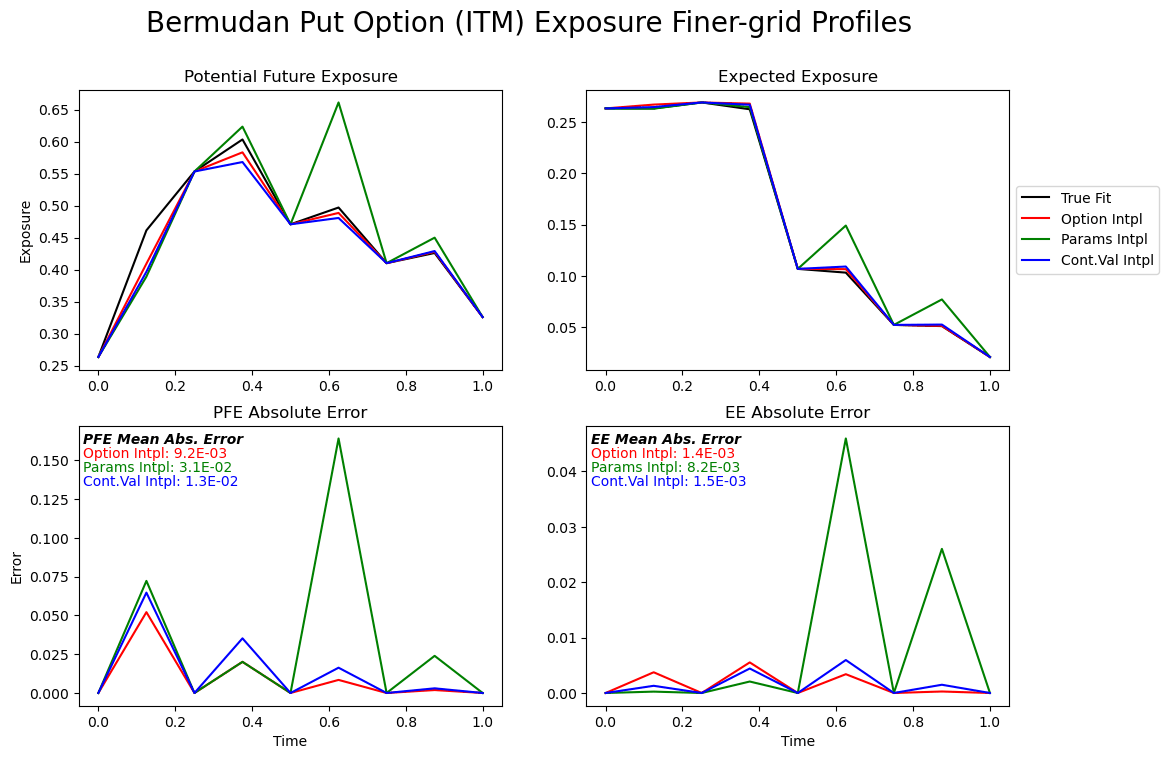}
\caption{\footnotesize The figure shows the exposure profiles of ITM Bermudan Option under risk-neutral measures at both exercise and finer grid horizons. In the first row, the left subplot corresponds to Potential Future Exposure, and the right plot corresponds to Expected Exposure. The exposures are generated by the Longstaff-Schwartz model using true fit (a cubic polynomial fit at finer horizons), option value interpolation (legend: Option Intpl), continuation value interpolation (legend: Cont.Val Intpl) and parameters interpolation (legend: Params Intpl). The second row highlights the corresponding interpolation errors with respect to the true model. The y-axis of the subplots in the first row corresponds to the exposure value. The y-axis of the subplots in the second row corresponds to the error value. The x-axis of all plots corresponds to time (in years).} 
\label{A2_bermudan_ITM_exposure_adj_finegrid_benchmark_longSchwarz.png}
\end{center}
\end{figure}

\begin{figure}[!htb]
\begin{center}
\includegraphics[width=0.8\textwidth, height=2.2in]{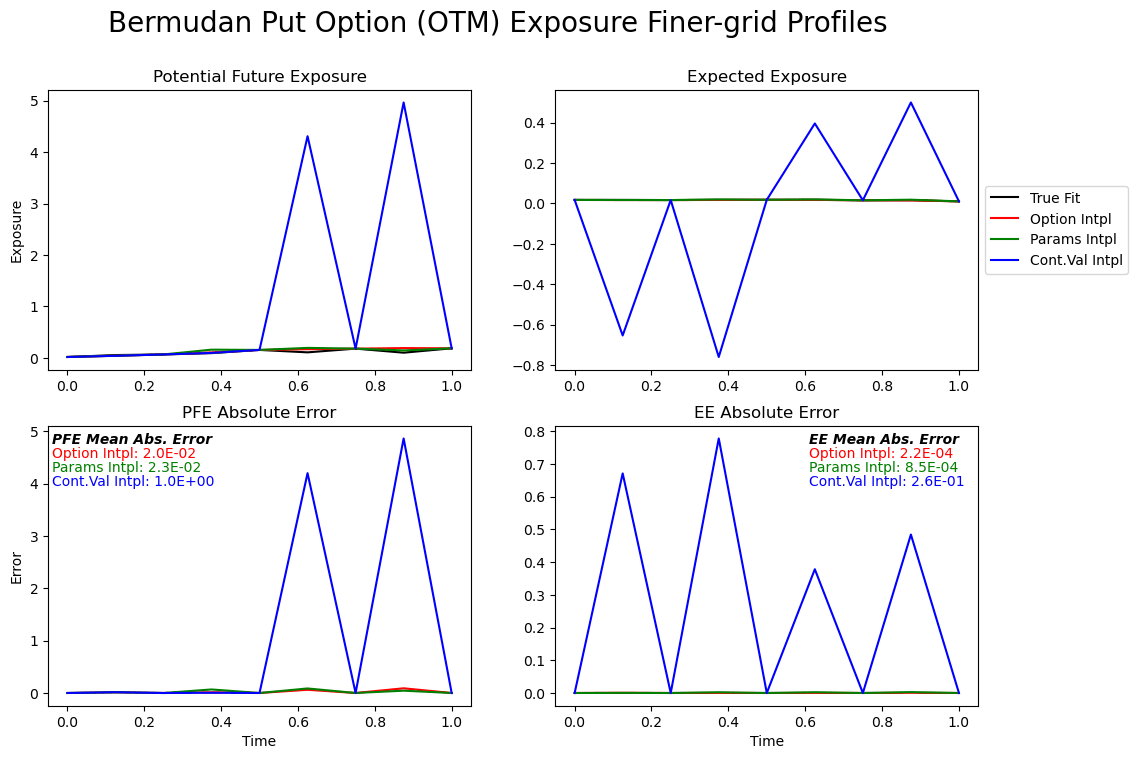}
\caption{\footnotesize The figure shows the exposure profiles of OTM Bermudan Option under risk-neutral measures at both exercise and finer grid horizons. In the first row, the left subplot corresponds to Potential Future Exposure, and the right plot corresponds to Expected Exposure. The exposures are generated by the Longstaff-Schwartz model using true fit (a cubic polynomial fit at finer horizons), option value interpolation (legend: Option Intpl), continuation value interpolation (legend: Cont.Val Intpl) and parameters interpolation (legend: Params Intpl). The second row highlights the corresponding interpolation errors with respect to the true model. The y-axis of the subplots in the first row corresponds to the exposure value. The y-axis of the subplots in the second row corresponds to the error value. The x-axis of all plots corresponds to time (in years).}
\label{A2_bermudan_OTM_exposure_adj_finegrid_benchmark_longSchwarz.png}
\end{center}
\end{figure}

\clearpage

\section{Longstaff Schwartz Algorithm}
\label{Longstaff Schwartz Algorithm}

We provide the summary of the Longstaff Schwartz algorithm with the cubic polynomial basis function used to compare the proposed model in this paper. Refer to Section \ref{Problem Formulation} for detailed information on the algorithm's notations. For detailed mathematical framework and convergence results, refer to the work by \citeauthor{longstaff2001valuing} (\citeyear{longstaff2001valuing}).\\

\noindent As already introduced in Section \ref{Problem Formulation}, we assume a complete probability space $(\Omega, \mathcal{F}, \mathbb{P})$,  filtration $\mathcal{F}_t: \ t \in [0,T]$ and an adapted underlying asset process $S_t$, $ \forall t$. The stochastic dynamics of the underlying asset are assumed to follow Geometric Brownian Motion (GBM), and therefore, 
 
\begin{equation}\label{GBM}
S_{t} = S_{0} \cdot  exp \left(  \Big( \ r - \frac{\sigma^2}{2} \ \Big) \ t + \sigma \ Z_{t} \right) , 
\end{equation} 

\noindent where, $S_{0}$ is the initial value of the underlying at time $0$, $r$ is the constant risk-free interest rate, $\sigma$ is the constant volatility and $Z_{t}$ is Brownian Motion. \\

\noindent We aim to price the target Bermudan option with strike $K$ $\in \mathbb{R}$ starting at time $t_0=0$ and expiring at $T$, with the right to exercise at $t_{m}$, where, $m \in \{1, 2, 3, \ldots, M \}$, and $t_{M} = T$.  \\ 

\noindent Let $h_t:= h(S_t)$ be an adapted process representing the option's intrinsic value; the holder of the option receives $max(h_t, 0)$ if the option is exercised at time $t$. $h_t = (S_t - K)$ for a call and $h_t = (K - S_t)$  for a put option. Assuming a risk-free savings account process, $B_t = exp(r \cdot t)$. \\

\noindent The Bermudan option price at $t_0$,
\begin{align}
\frac{V_{t_{0}} (S_{t_{0}})}{B_{t_{0}}} = \max_{\tau} \ \mathbb{E} \Bigg[\frac{h(S_{\tau})}{B_{\tau}}\Bigg] ,
\end{align}

\noindent where $V_{t}(.): \ t \in [0, \ T]$ is the option value function, and $\tau$ is the stopping time, taking values in the finite set $\{0, t_1, \ldots, T\}$. \\

\noindent The dynamic programming formulation to solve this optimisation problem is as follows. The value of the option at 
the terminal time $T$ is equal to the product's pay-off,

\begin{align}
V_T(S_T) = \max \big(h(S_T), 0\big).
\end{align}

\noindent Recursively, moving backwards in time, the following iteration is then solved, given $V_{t_m}$ has already been determined, the continuation or hold value $Q_{t_{m-1}}$ is given by:

\begin{align}
Q_{t_{m-1}} (S_{t_{m-1}}) = B_{t_{m-1}} \mathbb{E}\Bigg[  \frac{V_{t_m}(S_{t_m})}{B_{t_m}} | S_{t_{m-1}}  \Bigg].
\end{align}

\noindent At each $t_{m-1}$, $\forall m \in \{2, 3, \ldots, M\}$, $Q_{t_{m-1}} (S_{t_{m-1}}) $ is approximated by a cubic polynomial function $\zeta \Big(C(t_{m-1}), S_{t_{m-1}}\Big) $ as defined below: 

\begin{align}
\zeta \Big(C(t_{m-1}), S_{t_{m-1}}\Big) = c_0(t_{m-1}) + c_1(t_{m-1}) \cdot S_{t_{m-1}} \nonumber \\ 
+ c_2(t_{m-1}) \cdot (S_{t_{m-1}})^{2} + c_3(t_{m-1}) \cdot (S_{t_{m-1}})^{3},
\end{align}

\noindent where, $C(t_{m-1}) = [c_0(t_{m-1}), c_1(t_{m-1}), c_2(t_{m-1}), c_3(t_{m-1})]^{\intercal}$ are the polynomial coefficients, such that,

\begin{align}
 C(t_{m-1})  = \underset{C} {\mathrm{argmin}} \ \ L\Bigg[\zeta \Big(C, S_{t_{m-1}}\Big), \ B_{t_{m-1}} \cdot \frac{V_{t_{m}} \big(S_{t_{m}}\big)}{B_{t_m}}\Bigg],
\end{align}

\noindent and the loss function is defined as,

\begin{align}
&L\Bigg[\zeta \Big(C, S_{t_{m-1}}\Big), \ B_{t_{m-1}} \cdot \frac{V_{t_{m}} \big(S_{t_{m}}\big)}{B_{t_m}}\Bigg] := \nonumber \\
  &\sum_{j=1}^{N} \Bigg|\Bigg| \zeta (C, S_{t_{m-1}}(\omega_j)) - \ B_{t_{m-1}} \cdot \frac{V_{t_{m}} \big(S_{t_{m}}(\omega_j)\big)}{B_{t_m}} \Bigg|\Bigg|^{2}.
\end{align}


\noindent The bermudan option value at time $t_{m-1}$ for the underlying state $S_{t_{m-1}}$ is then given by:

\begin{align}
V_{t_{m-1}} (S_{t_{m-1}}) = \max \big( h(S_{t_{m-1}}), Q_{t_{m-1}} (S_{t_{m-1}}) \big), \\
\forall m \in \{2, 3, \ldots, M\}, \nonumber 
\end{align}

\noindent and the time zero price as,

\begin{align}
V_{t_0} (S_{t_0}) = \max \Bigg(h\Big(S_{t_0}\Big), \ \mathbb{E} \Bigg[ \frac{V_{t_1} (S_{t_1})}{B_{t_1}}  \Bigg] \Bigg). 
\end{align}

\clearpage

\section{COS Method for Bermudan Option Pricing}
\label{Appdx: COS Method for Bermudan Option Pricing}

\noindent In this paper, we assume a complete probability space $(\Omega, \mathcal{F}, \mathbb{P})$,  filtration $\mathcal{F}_t: \ t \in [0, T]$ and an adapted underlying asset process $S_t$, $ \forall t$. The stochastic dynamics of the underlying asset are assumed to follow Geometric Brownian Motion (GBM), and therefore, 
 
\begin{equation}\label{GBM}
S_{t} = S_{0} \cdot  exp \left(  \Big( \ r - \frac{\sigma^2}{2} \ \Big) \ t + \sigma \ Z_{t} \right) , 
\end{equation} 

\noindent where, $S_{0}$ is the initial value of the underlying at time $0$, $r$ is the constant risk-free interest rate, $\sigma$ is the constant volatility and $Z_{t}$ is Brownian Motion. \\

\noindent The COS method requires the knowledge of the characteristic function of the underlying asset price to value the options, and for GBM (with no dividend assumption on underlying asset price), it is given as,

\begin{align}
\phi^{GBM}(u, t) = e^{(i u \mu t - \frac{1}{2}\sigma^{2} u^{2}t)}, \nonumber \\
\mu := r - \frac{1}{2} \sigma^{2}.
\end{align}

\noindent We aim to price the target Bermudan option with strike $K$ $\in \mathbb{R}$ starting at time $t_0=0$ and expiring at $T$, with the right to exercise at $t_{m}$, where, $m \in \{1, 2, 3, \ldots, M \}$, $t_{M} = T$ and $t_{m} - t_{m-1} = $ $\Delta t$  $\forall m$. \\ 

\noindent Let $h_t:= h(S_t)$ be an adapted process representing the option's intrinsic value; the holder of the option receives $max(h_t, 0)$ if the option is exercised at time $t$. $h_t = (S_t - K)$ for a call and $h_t = (K - S_t)$  for a put option. Assuming a risk-free savings account process, $B_t = exp(r \cdot t)$. \\

\noindent The Bermudan option price at $t_0$,
\begin{align}
\frac{V_{t_{0}} (S_{t_{0}})}{B_{t_{0}}} = \max_{\tau} \ \mathbb{E} \Bigg[\frac{h(S_{\tau})}{B_{\tau}}\Bigg] ,
\end{align}

\noindent where $V_{t}(.): \ t \in [0, \ T]$ is the option value function, and $\tau$ is the stopping time, taking values in the finite set $\{0, t_1, \ldots, T\}$. \\

\noindent \citeauthor{fang2008novel} (\citeyear{fang2008novel}) leveraged the connection between Fourier coefficients of the probability density function and the characteristic function of the underlying asset price and introduced a novel pricing method called the COS method. For a detailed explanation of the relationship and its use in the COS method, refer \citeauthor{oosterlee2019mathematical} (\citeyear{oosterlee2019mathematical}). \citeauthor{fang2009pricing} (\citeyear{fang2009pricing}) extended this idea to derivatives with exotic features, with a specific emphasis on Bemrudan options, which is of key interest in the paper. \\

\noindent Let us define $x := ln \Big(\frac{S_{t_{m-1}}}{K}\Big)$ and $y := ln \Big(\frac{S_{t_m}}{K}\Big)$. Lets define the continuation value $ Q_{t}(x):= Q_{t}(S_{t})$, the value function $V_{t}(x) := V_{t}(S_{t})$ and the instrinsic value function $h_{t}(x):=  \max(h_{t}, 0)$ at time $t$. \\

\noindent The pricing formula for Bermudan option with $M$ exercise dates is given by, for each $m \in \{M, M-1, \ldots, 2\}$, 

\begin{align} \label{value functions}
Q_{t_{m-1}}(x) =  e^{-r \Delta t} \int_{\mathbb{R}} V_{t_m}(y) f(y|x) dy, \nonumber \\
V_{t_{m-1}}(x) = max(h_{t_{m-1}}(x), Q_{t_{m-1}}(x)),
\end{align}

\noindent and, 

\begin{align}
V_{t_0}(x) = e^{-r \Delta t} \int_{\mathbb{R}} V_{t_{1}}(x) f(y|x) dy,  \nonumber \\
V_{T}(x) = h_T(x) = \max \big( [\alpha K (e^{x} - 1)], 0 \big),
\end{align}

\noindent where, $\alpha=1$ for a call and $\alpha=-1$ for a put. \\ 

\noindent $f(y|x)$ is the probability density function of $y$ given $x$. The density function decays to zero rapidly as $y \rightarrow \pm \infty$. Therefore, without losing significant accuracy, the infinite integration of risk-neutral valuation is truncated to the interval $[a, b] \subset \mathbb{R}$. The cosine series expansion of $f(y|x)$ can be written as,

\begin{align} \label{f(y|x)}
f(y|x) = \sideset{}{'}\sum_{k=0}^{\infty}  A_{k}(x) cos \Big( k \pi \frac{y-a}{b-a} \Big), 
\end{align}

\noindent where the first summation term to be multiplied by $1/2$ denoted by the summation symbol $\sideset{}{'}\sum$. The series coefficients $\{A_k(x)\}_{k=0}^{\infty}$ can be defined by,

\begin{align}
A_k(x) := \frac{2}{b-a} \int_{a}^{b} f(y|x) cos \Big( k \pi \frac{y-a}{b-a} \Big) dy.
\end{align}

\noindent Substituting $f(y|x)$ from Equation \ref{f(y|x)} in Equation \ref{value functions} and interchanging the summation and integration operation yields,

\begin{align} \label{infinity sum of Q}
Q_{t_{m-1}}(x) = \frac{1}{2} (b-a) e^{-r \Delta t} \sideset{}{'}\sum_{0}^{\infty}  A_{k}(x) V_{k}(t_m),
\end{align}

and,

\begin{align} \label{series coefficient of value function}
V_{k} (t_m):= \frac{2}{b-a} \int_{a}^{b} V_{t_m}(y) cos \Big( k \pi \frac{y-a}{b-a} \Big) dy.
\end{align}

\noindent From Equation \ref{series coefficient of value function}, we could see that ${V_{k} (t_m)}_{k=0}^{\infty}$ are the Fourier-cosine series coefficients of the value function $V_{t_m} (y)$ on [a, b]. \\ 

\noindent The conditional characteristic function $\phi(u; x)$ is defined as,

\begin{align}
\phi(u; x) := \int_{\mathbb{R}}  f(y|x) e^{iuy} dy.
\end{align}

\noindent The fourier-cosine series coefficient $A_k(x)$ can be rewritten as,
\begin{align}
A_k(x) = \frac{2}{b-a} \Re \Bigg\{ e^{-ik\pi \frac{a}{b-a}}  \int_{a}^{b} e^{i \frac{k\pi}{b-a}y}  f(y|x) dy  \Bigg\},
\end{align}

\noindent where $\Re$ refers to the real part and, along with the following approximation of characteristic function, 
\begin{align}
\phi\Big( \frac{k\pi}{b-a} ; x \Big) &:= \int_{\mathbb{R}} e^{i \frac{k\pi}{b-a}y}  f(y|x) dy \nonumber \\
& \approx \int_{a}^{b} e^{i \frac{k\pi}{b-a}y}  f(y|x) dy
\end{align}

\noindent $A_k(x)$ can be approximated by $F_k(x)$ as,
\begin{align}
F_k(x) := \frac{2}{b-a} \Re \Bigg\{ \phi \Big(  \frac{k \pi}{b-a}; x \Big) e^{-ik\pi \frac{a}{b-a}} \Bigg\}, 
\end{align}

\noindent and for an exponential Levy process, it can be written as,
\begin{align}
F_k(x) := \frac{2}{b-a} \Re \Bigg\{ \phi \Big(  \frac{k \pi}{b-a} \Big) e^{-ik\pi \frac{a}{b-a}} \Bigg\}.
\end{align}

\noindent Further, by truncating the infinite series of the continuation value in Equation \ref{infinity sum of Q}, we get,

\begin{align}
Q_{t_{m-1}}(x) = e^{-r \Delta t}  \sideset{}{'}\sum_{k=0}^{L-1} \Re \Bigg\{  \phi \Big( \frac{k \pi}{b-a} \Big)  e^{ik\pi \frac{x-a}{b-a}} \Bigg\} V_k(t_m), 
\end{align}

\noindent and the time-zero price $V_{t_0}(x)$ can be approximated as,

\begin{align}
V_{t_0}(x) = e^{-r \Delta t} \sideset{}{'}\sum_{k=0}^{L-1} \Re \Bigg\{ \phi \Big( \frac{k \phi}{b-a} \Big) e^{i k \pi \frac{x-a}{b-a}} \Bigg\} V_k(t_1).
\end{align}

\noindent In the Equation \ref{series coefficient of value function}, the integral in the definition of Fourier cosine series coefficients of the value function can be split into two parts for two disjoint intervals of $x:  [a, x^{*}_m]$ and $(x^{*}_m, b]$ and the coefficients can be written for a call option as,
\begin{align}
V_k(t_m) = Q_{k}(a, x^{*}_{m}, t_m) + H_k(x^{*}_m, b),
\end{align}
\noindent and for a put option as,
\begin{align}
V_k(t_m) = H_k(a, x^{*}_{m}, t_m) + Q_{k}(x^{*}_m, b),
\end{align}

\noindent such that, $x^{*}_m$ is the point where continuation value equals intrinsic value called as early-exercise point, equivalently, $Q_{t_m}(x^{*}_m) = h_{t_m}(x^{*}_m)$. The Newton's method can be employed to determine $x^{*}_m$. Further, $H_k(x_1, x_2)$ and $Q_k(x_1, x_2, t_m)$ are the Fourier cosine series coefficients of the intrinsic value function and the continuation value function and can be determined analytically. \citeauthor{fang2009pricing} (\citeyear{fang2009pricing}) have presented an efficient algorithm for the computation of $\{V_k(t_m)\}$ based on the Fast Fourier Transform, which is used to price Bermudan options in this paper.

\end{appendices}


\bibliography{references}

\end{document}